\documentclass[preprint,notitlepage,superscriptaddress,letter]{revtex4-1}
\usepackage{amsmath}
\usepackage{amssymb}
\usepackage{graphicx}
\usepackage[caption=false]{subfig}
\usepackage{setspace}
\usepackage{txfonts}

\usepackage[T1]{fontenc}
\usepackage{textcomp}
\usepackage{lmodern}
\usepackage{color}

\renewcommand{\a}{\mathbf{a}}

\newcommand{\ND}[1]{\hat{n}_{#1}}
\renewcommand{\a}{\mathbf{a}}

\newcommand{\Ham}{\hat{H}}

\newcommand{\ketbra}[2]{\left|#1\middle>\middle<#2\right|}
\newcommand{\braket}[2]{\left<#1\middle|#2\right>}

\newcommand{\ket}[1]{\left|#1\right>}
\newcommand{\OPc}[2]{\hat{#1}_{#2}^{\dag}}
\newcommand{\OP}[2]{\hat{#1}_{#2}^{\vphantom{\dag}}}
\newcommand{\CD}[1]{\OPc{c}{#1}}
\newcommand{\C}[1]{\OP{c}{#1}}

\newcommand{\E}{\epsilon}

\newcommand{\expect}[1]{\left<#1\right>}

\newcommand{\BesselJ}{\mathcal{J}}
\newcommand{\Jn}[1]{\BesselJ_{#1}\left(A\right)}

\newcommand{\m}{\mathbf{m}}

\newcommand{\Ftwohop}{ \Lambda_{\m}^{(1)} }
\newcommand{\Fmidhop}{ \Lambda_{\m}^{(2)} }
\newcommand{\Fbackforth}{ \Gamma_{\m} }

\renewcommand{\S}{\mathbf{S}}

\newcommand{\thop}{t_{\textrm{h}}}

\newcommand{\mypara}[1]{\paragraph*{{#1}.---}}

\begin{document}
\title{Dynamical Time-Reversal Symmetry Breaking and Photo-Induced Chiral Spin Liquids in Frustrated Mott Insulators}
\date{\today}
\author{Martin Claassen}
 \affiliation{Department of Applied Physics, Stanford University, CA 94305, USA} 
 \affiliation{Stanford Institute for Materials and Energy Sciences, SLAC \& Stanford University, CA 94025, USA}
\author{Hong-Chen Jiang}
 \affiliation{Stanford Institute for Materials and Energy Sciences, SLAC \& Stanford University, CA 94025, USA}
\author{Brian Moritz}
 \affiliation{Stanford Institute for Materials and Energy Sciences, SLAC \& Stanford University, CA 94025, USA}
\author{Thomas P. Devereaux}
 \email[Author to whom correspondence should be addressed to: M. C. (\href{mailto:mclaassen@stanford.edu}{mclaassen@stanford.edu}) or T. P. D. (\href{mailto:tpd@stanford.edu}{tpd@stanford.edu})
]{}
\affiliation{Stanford Institute for Materials and Energy Sciences, SLAC \& Stanford University, CA 94025, USA}
\affiliation{Geballe Laboratory for Advanced Materials, Stanford University, Stanford, CA 94305, USA}

\begin{abstract}
The search for quantum spin liquids in frustrated quantum magnets recently has enjoyed a surge of interest, with various candidate materials under intense scrutiny. However, an experimental confirmation of a gapped topological spin liquid remains an open question. Here, we show that circularly-polarized light can provide a novel knob to drive frustrated Mott insulators into a chiral spin liquid (CSL), realizing an elusive quantum spin liquid with topological order. We find that the dynamics of a driven Kagome Mott insulator is well-captured by an effective Floquet spin model, with heating strongly suppressed, inducing a scalar spin chirality $\mathbf{S}_i \cdot (\mathbf{S}_j \times \mathbf{S}_k)$ term which dynamically breaks time-reversal while preserving SU(2) spin symmetry. We fingerprint the transient phase diagram and find a stable photo-induced CSL near the equilibrium state. The results presented suggest employing dynamical symmetry breaking to engineer quantum spin liquids and access elusive phase transitions that are not readily accessible in equilibrium.
\end{abstract}


\maketitle

Control of quantum materials out of equilibrium represents one of the grand challenges of modern condensed matter physics. While an understanding of general non-equilibrium settings beyond heating and thermalization is still in its infancy, a loophole concerns considering instead the transient quantum states of quasi-periodic perturbations such as wide-envelope laser pulses. Here, much of the intuition and language of equilibrium survives in a distinctly non-equilibrium setting within the framework of Floquet theory. While recently enjoying much attention and experimental success in the manipulation of single-particle spectra \cite{WangScience2013,KimScience2014,SieNMat2015,Mahmood2016} and band topology or short-range entangled topological states \cite{LindnerNatPhys2011,RudnerPRX2012,PoPRX2016}, a natural extension regards pumping of strongly-correlated systems. Here, the essence of Floquet physics lies not merely in imbuing one- \cite{WangScience2013} and two-particle \cite{KimScience2014} responses with the pump frequency as an additional energy scale, but in reshaping the underlying Hamiltonian to stabilize novel phases of matter that might be inaccessible in equilibrium.

Indeed, initial investigations suggest that the notion of effective low-energy physics persists in certain high-frequency regimes of time-periodic perturbations, leading for instance to enhancement of correlated hopping \cite{BukovPRL2016,BukovPRB2016}, strong-field sign reversal of nearest-neighbor Heisenberg exchange in a 1D magnet \cite{MentinkNComm2015,ItinPRL2015}, or enhancement of Cooper-pair formation \cite{KnapARXIV2015,SentefPRB2016,CoulthartARXIV2016}. Similar ideas are being pursued in the field of ultracold atoms to simulate artificial gauge fields, to dynamically realize topological band structures \cite{JotzuNature2014} or even propose a fractional quantum Hall effect in optical lattices \cite{YaoPRL2012,CooperPRL2013}. At the same time, recent advances in Floquet thermodynamics indicate that, while driven non-integrable closed systems are in principle expected to heat up to infinite temperature \cite{DAlessioPRX2014,LazaridesPRE2014}, heating can be exponentially slow on pre-thermalized time scales \cite{CanoviPRE2016,KuwaharaAnnPhys2016,MoriPRL2016,AbaninPRL2015,BukovPRL2015,HoARXIV2016} or altogether avoided via many-body localization \cite{DAlessioAnnPhys2013,PontePRL2015,LazaridesPRL2015} or dissipation \cite{dehghani2014dissipative,iadecola2015occupation,seetharam2015population}. An ideal condensed-matter realization hence entails a charge gap to limit absorption, as well as a delicate balance of competing phases, such that time-dependent perturbations and dynamical symmetry breaking can be expected to have an outsized effect and phase boundaries can be reached on pre-thermalized time scales with moderate effort.

Frustrated quantum magnets \cite{Balents2010} are prime candidates for such ideas. Strong local Coulomb repulsion between electrons freezes out the charge degrees of freedom, whereas the spin degrees of freedom are geometrically obstructed from ordering, hosting a delicate competition of conventionally-ordered phases as well as quantum spin liquids (QSLs) with long-range entanglement and exotic excitations \cite{Wen1990,WenPRB2002,NormanRMP2016,ZhouRMP2017,YamashitaScience2010,HanLeeNature2012,BanerjeeNMat2016,FuLeeScience2016}. The chiral spin liquid (CSL) constitutes one of the earliest proposals of a topologically-ordered QSL; it breaks time-reversal symmetry (TRS) and parity, while preserving SU(2) spin symmetry, and can be regarded as a bosonic $\nu=\textrm{\textonehalf}$ fractional quantum Hall state of spins with zero net magnetization and gapped semion excitations \cite{KalmeyerLaughlin1987,KalmeyerLaughlin1989,SchroeterPRL2007,ThomalePRB2009}. While an unlikely ground state in unperturbed microscopic models, recently the CSL was found to be a competing state \cite{BauerNComm2014,KumarPRB2015,WietekPRB2015,HickeyPRL2016,WietekARXIV2016,HePRL2014,GongSciRep2014,GongPRB2015}, in particular after explicit breaking of TRS and parity \cite{BauerNComm2014,KumarPRB2015,WietekPRB2015,HickeyPRL2016,WietekARXIV2016}. However, TRS breaking in experiment is realized canonically via external magnetic fields, necessarily entailing a Zeeman shift as the dominant contribution, which breaks SU(2) symmetry and disfavors CSLs \cite{BauerNComm2014}.

Here, we show that pumping a Mott insulator with circularly-polarized light below the Mott gap can dynamically break TRS without breaking of SU(2) or translation symmetry, providing a new knob to drive a frustrated quantum magnet into a CSL. Starting from a prototypical Hubbard model, the key questions posed by this work are three-fold: First, how does optically-induced time-reversal symmetry breaking manifest itself in a Mott insulator; second, can the ensuing effective Floquet spin model support a transient CSL and what are its signatures; and finally, does such an effective Floquet steady state description capture the many-body time evolution of an optically-driven Hubbard model? In the following, we answer all three questions affirmatively.

\section*{Results}

\mypara{Floquet-Hubbard Model} Our focus lies on Kagome antiferromagnets, which have recently garnered much attention due to novel candidate materials herbertsmithite, kapellasite and others \cite{NormanRMP2016}, with putative spin-liquid behavior at low temperatures. 
Experiments \cite{HanPRL2012} and first-principles calculations \cite{JansonPRL2008,JeschkePRB2013} indicate that the ground state and low-energy excitation spectra of these materials are well-captured by antiferromagnetic Heisenberg exchange between $d^9$ spins localized on Cu \cite{NormanRMP2016}. However, as photons couple to charge, a microscopic modelling of the light-matter interaction in principle must account for the multi-orbital structure at higher energies \cite{ClaassenNComm2016}, above the $\sim$2 eV charge gap \cite{HeltonPRL2007}. Here, we take a phenomenological approach, and, as an effective starting point that captures the essential physics but without pretense of a direct materials connection start from a driven single-orbital Hubbard model at half filling
\begin{align}
	\Ham(t) = -\thop \sum_{\left<ij\right>\sigma} e^{i \frac{e}{\hbar} \mathbf{r}_{ij} \cdot \mathbf{A}(t)}~ \CD{i\sigma}\C{j\sigma} + U \sum_i \ND{i\uparrow}\ND{i\downarrow} \label{eq:PumpedKagomeHubbard}
\end{align}
Here, $\thop,U,e$ denote nearest-neighbor hopping, Coulomb interaction and electron charge, $\mathbf{r}_{ij}$ denotes vectors between sites $i,j$, and $\mathbf{A}(t) = A(t) [ \cos(\Omega t), \sin(\Omega t) ]^{\textrm{T}}$ models a circularly-polarized pump beam with wide pulse envelope $A(t)$, coupling to electrons via Peierls substitution. Comparison of nearest-neighbor exchange $J \approx 4\thop^2/U$ with first-principles predictions for herbertsmithite \cite{JeschkePRB2013}
suggests $U/\thop$ of up to 40 due to the exceedingly narrow width of Cu $d$-orbital derived bands.

If $A(t)$ varies slowly with respect to the pump period, then the Hamiltonian becomes approximately periodic under a translation $\Ham(t + 2\pi/\Omega) = \Ham(t)$. Floquet theory then dictates that the behavior near the pump plateau is completely determined via many-body eigenstates of the form $\ket{\Psi_n(t)} = e^{-i \E_n t} \sum_m e^{im\Omega t} \ket{\Phi_m}$ with $\E_n$ the Floquet quasi-energy, where the $\ket{\Phi_m}$ conveniently follow as eigenstates of the static Floquet-Hubbard Hamiltonian
\begin{align}
	\Ham &= -\thop \sum_{\substack{\left<ij\right>\sigma \\ mm'}} \BesselJ_{m-m'}\left(A\right) e^{i (m-m') \arg \mathbf{r}_{ij}} \CD{i\sigma}\C{j\sigma} \otimes \ketbra{m}{m'} \notag\\
	&+  U \sum_{i} \ND{i\uparrow}\ND{i\downarrow} \otimes \mathbf{1} - \sum_{m} m\Omega ~\mathbf{1} \otimes \ketbra{m}{m} \label{eq:HamFloquetHubbard}
\end{align}
where $A$ denotes the dimensionless field strength at the pump plateau, such that $A(t) \approx A \hbar / (e a_0)$ with $a_0$ the nearest-neighbor distance, $m \in \mathbb{Z}$ is the Floquet index, and $\BesselJ_m(\cdot)$ denotes the Bessel function of the first kind [see Methods]. Note that the apparent Hilbert space expansion is merely a gauge redundancy of Floquet theory, as eigenstates with energy $\E_n + m\Omega$ identify with the same physical state $\forall m$. 

\mypara{Floquet Chiral Spin Model}   Physically, Eq. (\ref{eq:HamFloquetHubbard}) describes photon-assisted hopping in the presence of interactions, where electrons can enlist $m$ photons to hop at a reduced energy cost $U - m\Omega$ of doubly-occupying a site. Deep in the Mott phase the formation of local moments persists out of equilibrium as long as the pump remains off resonance and red-detuned from the charge gap. However, photon-assisted hopping reduces the energy cost of virtual exchange, pushing the system closer to the Mott transition and enlarging the range of virtual hopping paths that provide non-negligible contributions to longer-ranged exchange or multi-spin processes. Second, electrons acquire gauge-invariant phases when hopping around loops on the lattice for circular polarization. Crucially, and in contrast to an external magnetic field, an optical pump precludes a Zeeman shift, retaining the $SU(2)$ symmetry that is shared by CSL ground states. Symmetry considerations dictate that a manifestation of TRS breaking must to lowest-order necessarily involve a photo-induced scalar spin chirality $\chi_{ijk}$ term, with:
\begin{align}
	\Ham_{\textrm{spin}} = \sum_{ij} J_{ij}~ \S_i \cdot \S_j + \sum_{ijk} \chi_{ijk}~ \S_i \cdot \left( \S_j \times \S_k \right)   \label{eq:HeisenbergChiralHam}
\end{align}

\begin{figure}
	\centering
	\includegraphics[width=12cm]{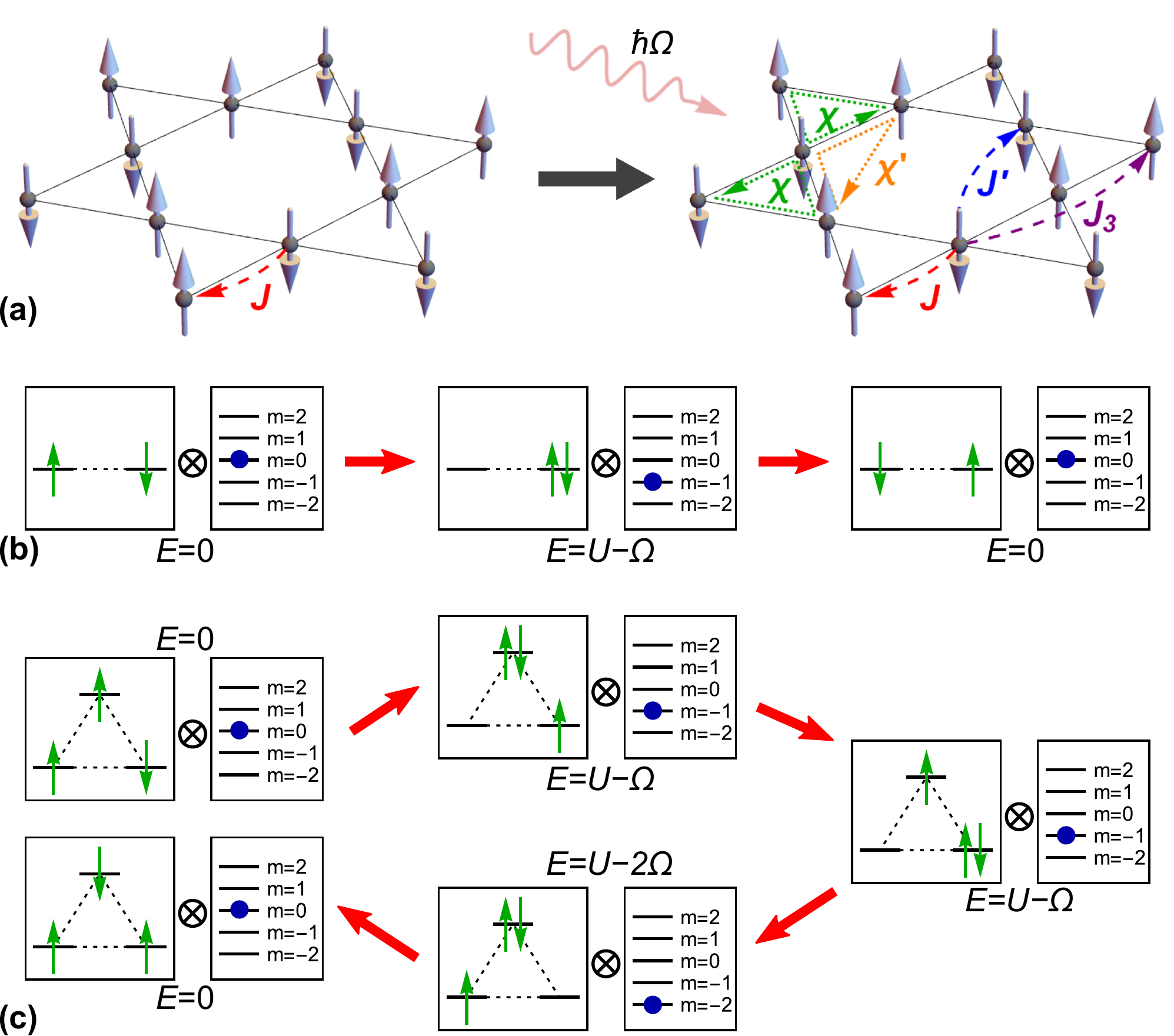}
	\caption{\textbf{Photo-induced Kagome chiral antiferromagnet.} (a) Starting from a Kagome Mott antiferromagnet, pumping with circularly-polarized light dynamically breaks time-reversal and parity while preserving SU(2) spin symmetry, photo-inducing scalar spin chirality $\S_i \cdot (\S_j \times \S_k)$ contributions on elementary equilateral ($\chi$) and isosceles  ($\chi'$) triangles. Pump strength and frequency provide knobs to tune $\chi,\chi'$ as well as Heisenberg exchange $J,J',J_3$, as described in the main text. Examples of (b) nearest-neighbor and (c) three-site Floquet virtual hopping processes including absorption of photons, in the Mott-insulating regime. Boxes graphically depict the example initial, virtual intermediate and final states for second- and fourth-order virtual hopping processes, in terms of the product space of electronic degrees of freedom and the Floquet index. TRS is broken in (c), inducing scalar spin chiralities $\chi$ on triangles of the lattice, whereas (b) solely induces nearest-neighbor Heisenberg exchange.}
	\label{fig:couplings}
\end{figure}

This Floquet Chiral Spin Hamiltonian is the central focus of the paper; to derive it microscopically from the driven Kagome-Hubbard model (\ref{eq:PumpedKagomeHubbard}), it is instructive to first consider the high-frequency limit $\Omega \gg U, \thop$. Here, circularly-polarized pumping induces complex nearest-neighbor hoppings $\tilde{t} = \thop(1-A^2/4) + i(\sqrt{3}/4) \thop^2 A^2/\Omega$ as well as purely-complex next-nearest-neighbor hoppings $\tilde{t}' = -i(\sqrt{3}/4) \thop^2 A^2/\Omega$, analogous to a staggered magnetic flux pattern in the unit cell [see Supplementary Note 1]. To third order in $\tilde{t},\tilde{t}'$, a spin description then includes scalar spin chirality contributions, with $\chi = 9 \sqrt{3} \thop^4 A^2 / 2 U^2 \Omega$ of equal handedness for both equilateral triangles per unit cell, as depicted in Fig. \ref{fig:couplings}(a), and six isosceles triangles of opposite handedness with $\chi' = \chi/3$, such that the total chiral couplings in the unit cell sum to zero.

Now consider sub-gap pumping $\Omega < U$. Starting from Eq. (\ref{eq:HamFloquetHubbard}), a microscopic derivation of the Floquet spin Hamiltonian proceeds via quasi-degenerate perturbation theory, where care must be taken to simultaneously integrate out $m\neq 0$ Floquet states and many-body states with doubly-occupied sites [see Supplementary Note 2]. Fig. \ref{fig:couplings}(b) and (c) depict relevant virtual processes, involving simultaneous hopping of electrons and absorption of $m$ photons with intermediate energy cost $U-m\Omega$. To second order in virtual hopping, two-site exchange processes [Fig. \ref{fig:couplings}(b)] are phase-agnostic and solely renormalize the nearest-neighbor Heisenberg exchange $\tilde{J} = 4 \sum_m |\Jn{m}|^2 \thop^2 / (U - m\Omega)$ \cite{MentinkNComm2015,BukovPRL2016}. While every process contributes to Heisenberg exchange, a scalar spin chirality contribution appears for multi-hop processes that enclose an area. Na\"ively, to third order, an electron could simply circumnavigate the elementary triangles of the Kagome lattice; however, these processes interfere destructively and cancel exactly to all orders in $A$ even though time-reversal symmetry is broken, and in contrast to an external magnetic field [see Supplementary Note 2]. This is consistent with results on the resonant A$_{\textrm{2g}}$ Raman response of Mott insulators \cite{KoPRB2010}, that connect to the $A \to 0, m=1$ limit. Instead, time-reversal symmetry breaking first manifests itself to fourth order in virtual hopping. Here, processes [Fig. \ref{fig:couplings}(c)] can either encompass an elementary triangle, or virtually move an electron back and forth two legs of a hexagon, inducing scalar spin chirality contributions as shown in Fig. \ref{fig:couplings}(a), with
\begin{align}
	\chi &= 3 \sum_{\m} \left[ \sin\left[\tfrac{2\pi(m_1-m_2+m_3)}{3}\right] \Ftwohop - \sin\left[\tfrac{2 \pi m_2}{3}\right] \Fmidhop \right] \label{eq:ChiEff} \\
	\chi' &= \sum_{\m} \left[ \sin\left[\tfrac{\pi(m_1-m_2+m_3)}{3}\right] \Ftwohop - \sin\left[\tfrac{\pi m_2}{3}\right] \Fmidhop \right]
\end{align}
where $\m = \{m_1,m_2,m_3\}$ are Floquet indices, and
\begin{align}
	\Ftwohop &= \frac{8 \thop^4 \Jn{m_1} \Jn{m_2-m_1} \Jn{m_2-m_3} \Jn{m_3} }{\prod_{i=1}^3 (U-m_i\Omega)} \\
	\Fmidhop &=  2(1-\delta_{m_2}) \tfrac{U-m_2\Omega}{m_2\Omega} (-1)^{m_1-m_3} \cos^2\left[\tfrac{m_2 \pi}{2}\right] \Ftwohop
\end{align}
Here, $\Ftwohop$ and $\Fmidhop$ parameterize fourth-order virtual hopping processes for which the second intermediate virtual state retains a single double-occupied site or returns to local half filling (albeit with non-zero Floquet index), respectively.
Furthermore, next-nearest-neighbor Heisenberg exchange
\begin{align}
	J' &= \sum_{\m} \left[ \Fbackforth - \cos\left[\tfrac{\pi(m_1-m_2+m_3)}{3}\right]  \frac{\Ftwohop}{2} + \cos\left[\tfrac{\pi m_2}{3}\right] \frac{\Fmidhop}{2} \right]  \notag\\
	J_3 &= \sum_{\m} \left[ \Fbackforth - \frac{\Ftwohop}{2}  + \frac{\Fmidhop}{2}  \right]
\end{align}
and corrections to nearest-neighbor Heisenberg exchange
\begin{align}
	J &= \tilde{J} + \sum_{\m} \left\{ \left[ 2 + \cos\left[\tfrac{\pi (m_1-m_3)}{3}\right] + \cos\left[\tfrac{2\pi (m_1-m_3)}{3}\right] +  \right.\right. \notag\\
	& \left. + \cos\left[\tfrac{\pi (m_1-m_2+m_3)}{3}\right] + \tfrac{1}{2} \cos\left[\tfrac{2\pi (m_1-m_2+m_3)}{3}\right] \right]  \Ftwohop \notag\\
	&- \left[  1 + 2 \cos\left[\pi m_2\right] + \cos\left[\tfrac{\pi m_2}{3}\right] + \tfrac{1}{2} \cos\left[\tfrac{2\pi m_2}{3}\right]   \right] \Fmidhop - 7 \Fbackforth \left.\vphantom{\tfrac{1}{2}}\right\}
\end{align}
appear at the same order, with
\begin{align}
	\Fbackforth &= \left[ \frac{\delta_{m_3,0} }{U-m_1\Omega} + \frac{\delta_{m_3,0} }{U-m_2\Omega} \right] \prod_{i=1}^2 \frac{2\thop^2 \left[\Jn{m_i}\right]^2}{(U-m_i\Omega)}  \label{eq:LastEffParam}
\end{align}
parameterizing a two-fold virtual nearest-neighbor exchange process.

\begin{figure}
	\includegraphics[width=12cm]{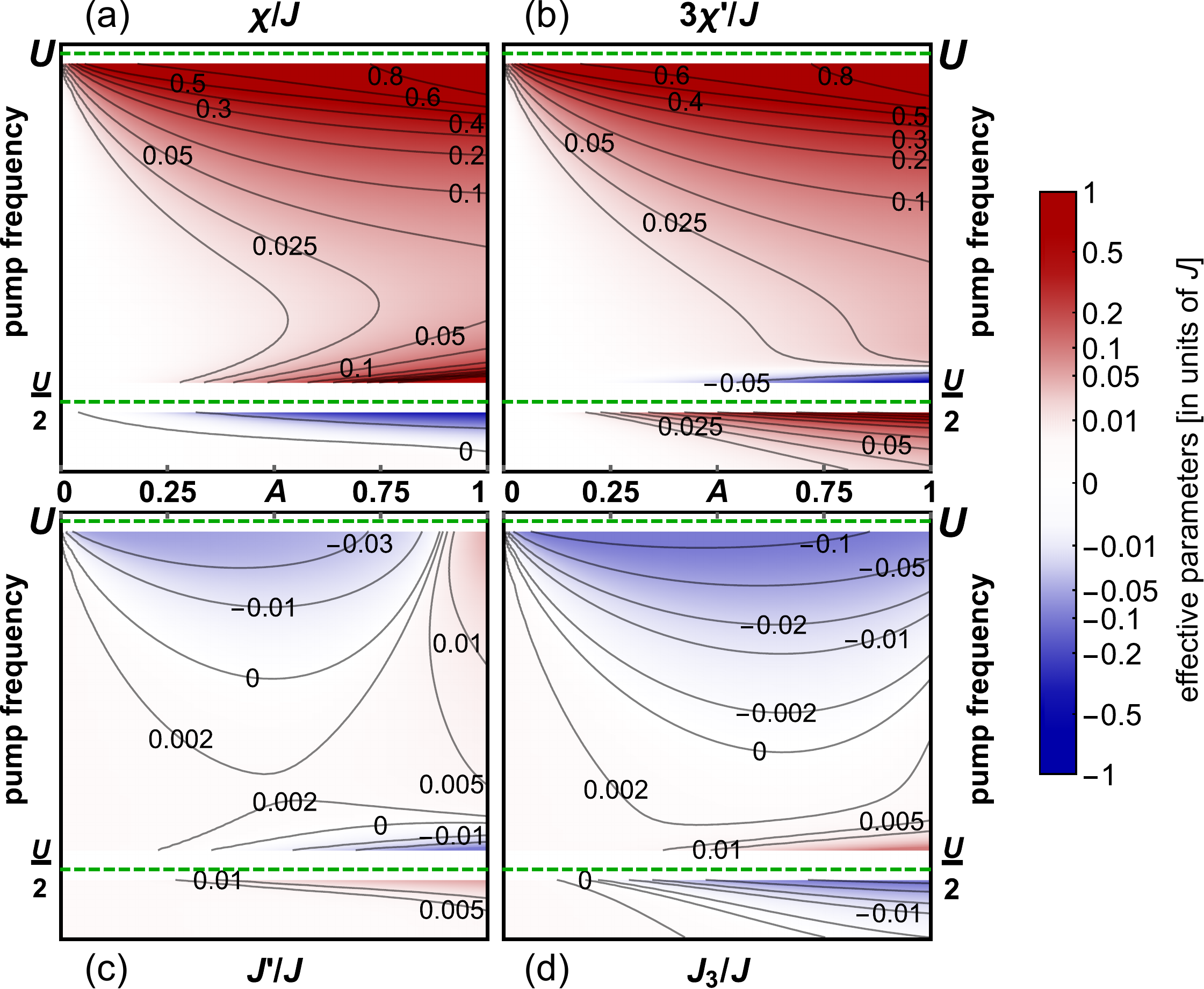}
	\caption{\textbf{Pump dependence of chiral spin model parameters.} Photo-induced time reversal symmetry (TRS) breaking scalar spin chirality $\chi$, $\chi'$ [(a), (b)] and TRS preserving Heisenberg exchange $J'$, $J_3$ [(c), (d)] terms for a Kagome Mott insulator pumped by circularly-polarized light, as a function of pump strength $A$ and pump frequency [from Eqns. (\ref{eq:ChiEff}-\ref{eq:LastEffParam})]. Parameters are depicted in units of nearest-neighbor exchange coupling $J$. Dashed lines indicate one- and two-photon resonances with respect to the Hubbard repulsion $U$. Equal signs of $\chi,\chi'$ signify a staggering of $\S_i \cdot (\S_j \times \S_k)$ contributions between elementary triangles and isosceles triangles inside the hexagons.}
	\label{fig:HeisenbergChiralTerms}
\end{figure}

\mypara{Steady-State Phase Diagram} Having established the effective steady-state physics for the duration of the pump pulse, the next question concerns whether the photo-induced Floquet spin model [Eq. \ref{eq:HeisenbergChiralHam}] can indeed stabilize a CSL. Consider its parameter space as a function of $A,\Omega$, depicted in Fig. \ref{fig:HeisenbergChiralTerms} for $U=20\thop$. Adiabatic ramping up of the circularly-polarized pump then corresponds to horizontal trajectories with fixed $\Omega$. Fig. \ref{fig:HeisenbergChiralTerms} (a) and (b) show that TRS-breaking scalar spin chiralities develop with increasing field strength, whereas the effect on longer-ranged Heisenberg exchange [(c) and (d)] is comparatively weak. This immediately suggests that circularly-polarized pumping grants a handle to change the underlying many-body state. Close to the one-photon resonance ($\Omega = U$), chiral contributions are staggered between elementary and isosceles triangles [Fig. \ref{fig:couplings}(a)], whereas $\chi'$ changes sign when $\Omega$ approaches a two-photon resonance ($\Omega = U/2$).

We analyze the steady state phase diagram using exact diagonalization of the Floquet spin model [Eq. \ref{eq:HeisenbergChiralHam}], parameterized by pump strength and frequency. In equilibrium ($A=0$), the ground state of Eq. (\ref{eq:HeisenbergChiralHam}) is gapped and TRS invariant. Absence of conventional spin order is evidenced by a rapid decay of spin-spin and chiral-chiral correlation functions on a 36-site cluster [Fig. \ref{fig:CSL}(a)], consistent with density-matrix renormalization group simulations that find a gapped $Z_2$ QSL \cite{JiangPRL2008, YanScience2011, JiangNPhys2012, DepenbrockPRL2012}. We adopt this view for the thermodynamic limit, but note that the ground state degeneracy of a $Z_2$ QSL remains inaccessible in exact diagonalization of finite-size clusters [see Methods]. Upon pumping ($A \neq 0$), the spin correlator displays no propensity for ordering; however, chiral correlations develop smoothly [Fig. \ref{fig:CSL}(a)]. Importantly, a two-fold ground state quasi-degeneracy develops continuously, with a gap to many-body excitations, indicative of a CSL. To track the phase boundary as a function of $A,\Omega$, we determine the parameter space region within which the ground state degeneracy as well as the gap $\Delta_{\textrm{CSL}}$ to the many-body excitation manifold above the CSL survives insertion of a flux quantum through the torus [see Methods]. As shown in Fig. \ref{fig:CSL}(b), a robust photo-induced CSL develops already for weak $A$, with excited states well-separated in energy. Finally, a proper verification of the CSL necessitates characterizing its ground state topological order. We therefore fingerprint the photo-induced phase by determining a basis of minimally-entangled states from combinations of the two degenerate ground states $\ket{\psi_{1,2}}$, minimizing the Renyi entropy for their reduced density matrices in two distinct bipartitions [Fig. \ref{fig:CSL}(d)] on a 36-site torus [see Methods]. $C_6$ symmetry allows extraction of both modular $\mathcal{U}$, $\mathcal{S}$ matrices \cite{ZhangPRB2012,CincioPRL2013,ZhuPRB2013} that encode self- and mutual-braiding statistics of the elementary excitations. We find that the photo-induced CSL corresponds uniquely to a $\nu=\tfrac{1}{2}$ bosonic fractional quantum Hall state, with matching modular matrices \cite{Wen1990}
\begin{align}
	\mathcal{U} = e^{-i \frac{2\pi}{24} \times 1.02} \left[\begin{array}{cc} 1 & 0 \\ 0 & 0.99 i \end{array}\right], ~  	\mathcal{S} = \tfrac{1}{\sqrt{2}} \left[\begin{array}{cc}  1.00 & 1.00 \\ 1.00 & -1.02 \end{array}\right]
\end{align}


\newcommand{\tFast}{t}
\newcommand{\tSlow}{T}

\mypara{Time Evolution} Having established a photo-induced CSL for the Floquet spin model, the final question concerns whether this effective spin model qualitatively captures the time evolution of the driven Hubbard model [Eq. \ref{eq:PumpedKagomeHubbard}]. 

To this end, we consider a circularly-polarized optical pump pulse with a slow sinusoidal ramp-up and a wide pump plateau [Fig. \ref{fig:HubbardVsFloquet}(a)], and simulate the exact many-body dynamics of driven 12-site $U=30$ Kagome Hubbard clusters for long times $t \leq 1000~\thop^{-1}$. Conceptually, the transient state can then be thought of as dynamically following the instantaneous Floquet eigenstate $\ket{\Psi(\tau,\bar{T}_{\textrm{slow}})} \approx e^{-i \E(\bar{T}_{\textrm{slow}}) \tau} \sum_m e^{i m\Omega \tau} \ket{\Phi(\bar{T}_{\textrm{slow}})}$, with the time variable $t$ ``separating'' into fast ($\tau$) and slow ($\bar{T}_{\textrm{slow}}$) moving components. Reaching the pump plateau, the time-evolved state will nevertheless retain a finite quasi-energy spread $\rho(\E_\alpha)$ with $\ket{\Psi(t)} = \sum_\alpha \rho(\E_\alpha) e^{-i \E_\alpha t} \sum_m e^{i m\Omega t} \ket{\Phi_\alpha(t)}$ (with $\alpha$ indexing the Floquet eigenstates). While dephasing of these constituent Floquet eigenstates should ultimately thermalize the system to infinite temperature, the system nevertheless matches the effective spin dynamics described by Eq. \ref{eq:HeisenbergChiralHam}, and barely absorbs energy on the broad ``pre-thermalized'' time scales of interest, as we show below.

First, to compare to the Floquet spin description, we focus on time-dependent scalar spin chirality expectation values $\chi_{ijk}(t) = \expect{ \S_i \cdot (\S_j \times \S_k) }$ ($\S_i = \CD{i\sigma} \vec{\mathbf{s}}_{\sigma\sigma'} \C{i\sigma'}$) on elementary triangles of the Kagome cluster [Fig. \ref{fig:HubbardVsFloquet}(b)]. Vanishing in equilibrium due to TRS, the pump-period average of $\chi_{ijk}(t)$ should saturate to its Floquet expectation value at the pump plateau. Fig. \ref{fig:HubbardVsFloquet}(c) compares $\chi_{ijk}(t)$, time-averaged over the pump plateau, to corresponding static $\chi_{ijk}$ expectation values of the Floquet spin model (\ref{eq:HeisenbergChiralHam}) ground state. The latter follows from choosing $\chi,\chi',J,J',J_3$ via Eqns. (\ref{eq:ChiEff})-(\ref{eq:LastEffParam}), with $A,\Omega$ the pump parameters of the Hubbard time evolution. Intriguingly, the electronic time evolution is in excellent qualitative agreement with predictions for the Floquet spin model, even when driven close to the Mott transition.

Quantitative discrepancies predominantly originate from deviations of the local moment $\expect{(S^z)^2} < 1/4$; additionally, the 12-site cluster with periodic boundary conditions permits weak ring exchange contributions from loops of virtual hopping around the cluster. Importantly, the transient increase in double occupancies is not an indication of heating -- instead, this follows from a reduction of the effective $U$ of the transient Floquet-Hubbard Hamiltonian, a consequence of the photo-assisted hopping processes depicted in Fig. \ref{fig:couplings} (b) and (c). Quantitative differences between spin and fermionic observables are therefore  analogous to differences between canonical spin and fermionic descriptions of equilibrium quantum magnets for a finite Hubbard-$U$.

To analyze this in detail, we focus on pumping the system across the charge resonance with the upper Hubbard band, where the photo-induced scalar spin chirality contribution is expected to be largest. Figs. \ref{fig:heating}(a)-(c) show the period-averaged double occupancy $\expect{\ND{\uparrow}\ND{\downarrow}}$ as a function of pump strength and detuning from the charge resonance $\approx U - 5.5 \thop$. Upon resonant charge excitation, the system heats up rapidly and the double occupancy approaches its infinite-temperature limit $\expect{\ND{\uparrow}\ND{\downarrow}} \to 1/4$. Importantly, this entails that thermalization at long times is independent of the pump strength $A$.

Conversely, in the off-resonant regime, one observes a pump-strength dependent saturation of the double occupancy. Here, proper heating is strongly suppressed and the system instead realizes the effective Floquet chiral quantum magnet with a transient reduction of the Hubbard interaction $U$. To verify that the driven steady state indeed follows the ground state of the effective Floquet Hamiltonian adiabatically, consider a period-shifted Floquet ``fidelity measure'' $\mathcal{F} = \left|\braket{\Psi(t+T)}{\Psi(t)}\right|$, where $T = 2\pi/\Omega$ is the pump period. At the pump plateau with discrete time translation symmetry, $\mathcal{F}$ is time-independent and quantifies the Floquet quasi-energy spread of the transient steady state [Fig. \ref{fig:heating}(d)]. For a pure Floquet state, $1 - \mathcal{F} \to 0$, suggesting that the driven state below resonance adiabatically follows a Floquet eigenstate, whereas adiabaticity is lost when crossing the absorption edge.

To distinguish residual heating on these pre-thermalized time scales of interest \cite{AbaninPRL2015,HoARXIV2016} from a transient increase in energy in the chiral quantum magnet due to modulation of the Hamiltonian, consider the period-averaged stroboscopic energy operator $\hat{\expect{E}}$ [see Methods]. On the pump plateau, both the double occupancy and $\hat{\expect{E}}$ saturate to their pre-thermalized steady state expectation values; however, a minuscule residual gradient over thousands of pump cycles remains. To good approximation for the time scales considered here, we can linearize the energy on the pump plateau $\hat{\expect{E}}(t) \approx E_0 + t \Delta E$, and extract the heating rate $\Delta E$ from simulations. Fig. \ref{fig:heating}(e) depicts the absorbed energy per pump cycle on the pump plateau, as a function of pump strength and detuning from the absorption edge. Remarkably, residual heating is largely suppressed close to resonance, with an absorbed energy on the order of $10^{-6}\thop$ per pump cycle. Na\"ively, this extraordinary meta-stability suggests that it could take on the order of tens of thousands of pump cycles for heating to dominate the dynamics, absorbing a total energy $\sim J$ the exchange coupling.

A more microscopic analysis of photo-excitation for realistic materials will likely lead to a less optimistic upper bound on the time scales of interest. First, a materials-specific modelling of electron-photon coupling and multi-band effects will modify the effective photo-induced spin Hamiltonian, albeit necessarily retaining the salient symmetry properties and scalar spin chirality contributions that stabilize the CSL. Second, an intriguing follow-up question regards the role of coupling to - and heating of - the lattice. While magneto-elastic coupling to phonons is weak in most materials and the optical frequencies under consideration are far from resonance with infrared-active phonon modes, electron phonon coupling will nevertheless indirectly heat the lattice due to Raman-assisted hopping processes. Conversely, the separation of time scales for electrons and phonons suggests that the phonon bath could similarly play the role of a dissipative channel, effectively ``cooling'' the electronic system. While initial investigations have already studied the case of free or weakly-interacting electrons \cite{dehghani2014dissipative,iadecola2015occupation,seetharam2015population}, a proper understanding of the confluence of strong interactions, external drive and dissipation remains an interesting topic for future study.

\section*{Conclusions}
	
In summary, we have shown that pumping a frustrated Mott insulator with circularly-polarized light can dynamically break TRS while preserving SU(2) symmetry of the underlying spin system, by augmenting its effective dynamics with a transient scalar spin chirality term. Remarkably, on the Kagome lattice this effective Floquet spin model was found to stabilize a transient CSL in a broad parameter regime. Our results suggest that wide-pulse optical perturbations can provide an intriguing knob to tune the low-energy physics of frustrated quantum magnets, shedding light on novel regions of their phase diagram hithero unexplored.


\acknowledgements{
We acknowledge support from the U. S. Department of Energy, Office of Basic Energy Sciences, Division of Materials Sciences and Engineering under Contract No. DE-AC02-76SF00515. Computational resources were provided by the National Energy Research Scientific Computing Center supported by the Department of Energy, Office of Science, under Contract No. DE- AC02-05CH11231.
}

\section*{Author Contributions}

M. C. conceived the project, developed the theoretical description and performed the numerical simulations. M. C., H.-C. J. and B. M. analyzed the numerical results. The manuscript was written by M. C. with input from all authors. T. P. D. supervised the project.

\section*{Competing Financial Interests}

The authors declare no competing financial interests.

\section*{Methods}

\subsection*{Floquet Theory}

Consider a generic time-dependent many-body Hamiltonian with discrete time-translation invariance $\Ham(t) = \Ham(t + 2\pi/\Omega)$. Instead of solving the many-body time evolution, one can reexpress the time-dependent Schr\"odinger equation in a Floquet eigenbasis $\ket{\Phi_\alpha(t)} = e^{-i\epsilon_\alpha t} \sum_m e^{im\Omega t} \ket{u_{\alpha,m}}$, where $\alpha$ indexes the basis wave function and $\epsilon_\alpha$ is its respective Floquet quasi-energy. Then, determination of the time-dependent eigenstates of the driven system reduces to finding the time-independent eigenstates of the Floquet Hamiltonian
\begin{align}
	\Ham_{\textrm{floquet}} = \sum_{mm'} \left[ \Ham_{m-m'} - m\Omega \delta_{m,m'} \right] \ketbra{m}{m'}
\end{align}
where $\Ham_{m-m'}$ are the Fourier expansion coefficients of $\Ham(t)$. Taking $\Ham(t)$ as the driven Kagome-Hubbard model of Eq. (\ref{eq:PumpedKagomeHubbard}) in the main text straightforwardly recovers the Floquet-Hubbard Hamiltonian [Eq. (\ref{eq:HamFloquetHubbard})]. Here, the dimensionless pump strength $A$ that enters via Peierls substitution, relates to the electric field $\mathcal{E}$ as $\mathcal{E} = \Omega / (e a_0)$, with $e$ the electron charge and $a_0$ the nearest-neighbor bond distance. While realistic estimates for materials would necessarily entail significant contributions from multi-orbital effects and local dipole transitions, all of which are not captured within the single-orbital Hubbard model, a na\"ive estimate from solely Peierls substitution for a single-orbital approximation of herbertsmithite yields $A=0.25 \dots 0.5$ for $\mathcal{E} \dots 100 \dots 200 \textrm{meV}\AA^{-1}$ for a 900nm near-infrared pump.

\subsection*{Numerical Simulations}

To characterize the photo-induced chiral spin liquid state, we performed exact diagonalization calculations of the Floquet chiral spin Hamiltonian, described in Eq. (\ref{eq:HeisenbergChiralHam}). Due to three-spin interactions and longer-ranged exchange interactions, the sparsity of the resulting Hamiltonian matrix is over an order of magnitude lower than for a nearest-neighbor Heisenberg antiferromagnet. Spin and chiral correlation functions as well as minimally-entangled states were calculated for a 36-site cluster with periodic boundary conditions, spanned by vectors $\mathbf{R}_1 = 4\mathbf{a}_1 - 2\mathbf{a}_2, ~\mathbf{R}_2 = -2\mathbf{a}_1 + 4\mathbf{a}_2$, where $\mathbf{a}_1, \mathbf{a}_2$ are the lattice vectors. This choice retains the rotational symmetry of the Kagome lattice, facilitating extraction of the modular matrices. Previous exact diagonalization studies have simulated the 36-site cluster for a purely nearest-neighbor Heisenberg model \cite{LaeuchliPRB2011,NaganoJPhys2011}. While density-matrix renormalization group studies indicate that such a model is likely to stabilize a $Z_2$ QSL in the thermodynamic limit, it is well-known that its ground state degeneracy remains inaccessible in exact diagonalization of finite-size clusters \cite{LaeuchliPRB2011,NaganoJPhys2011}, and the 36-site cluster is likely to host a QSL close to a phase boundary with a valence bond crystal \cite{LiARXIV2011}. However, the nature of the equilibrium ground state and its extrapolation to the thermodynamic limit does not affect the conclusions regarding transient state; crucially, the chiral spin liquid state is well-stabilized already on the finite-size systems under consideration, with a robust many-body excitation gap.

\begin{figure}
	\centering
	\includegraphics[width=16cm]{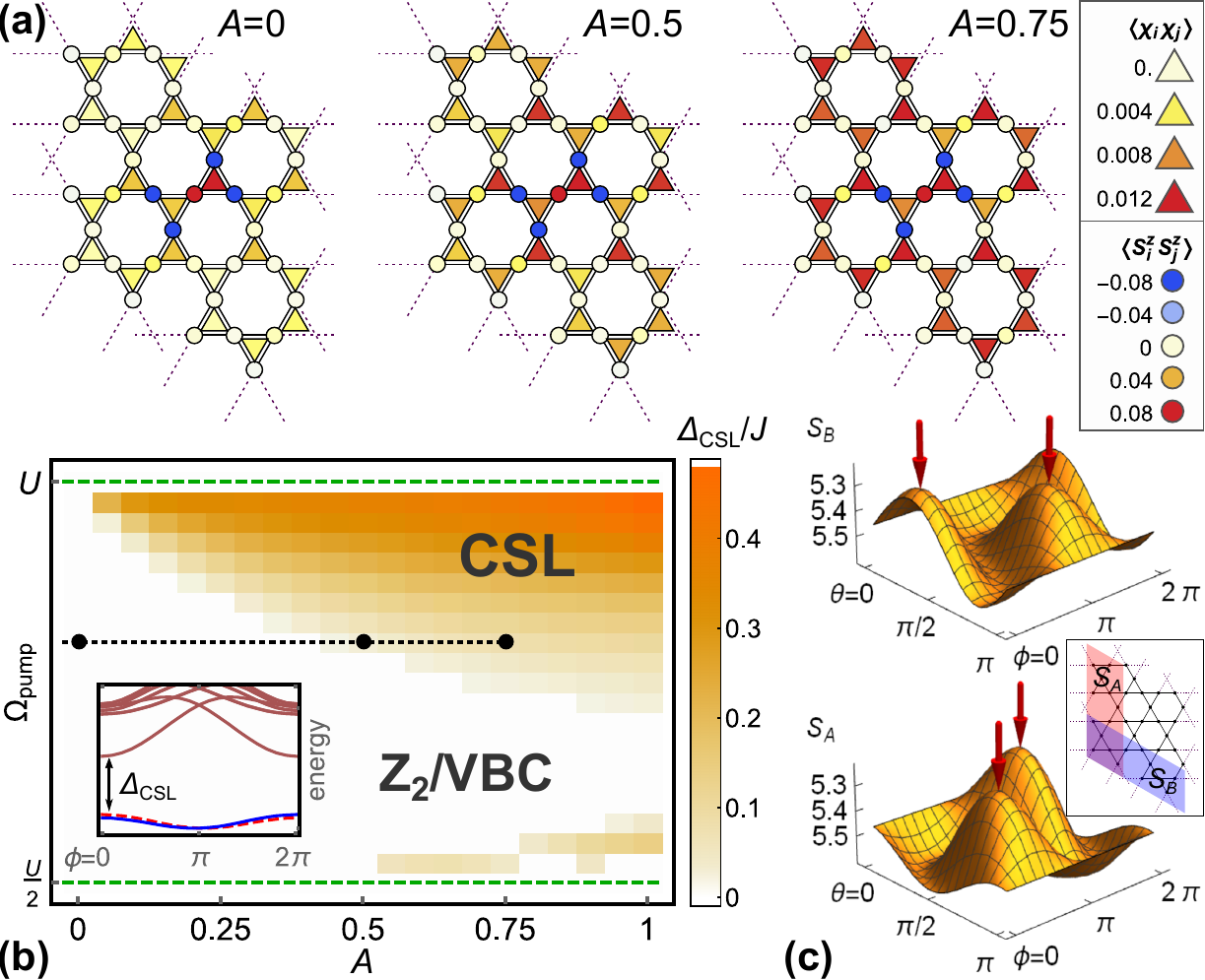}
	\caption{\textbf{Identification of a photo-induced chiral spin liquid.} (a) Evolution of spin $\expect{\S_i \cdot \S_j}$ (circles on vertices) and chiral $\expect{\left[ \S_i \cdot (\S_j \times \S_k) \right] \left[ \S_l \cdot (\S_m \times \S_n) \right]}$ (triangles) correlation functions as a function of pump strength $A$, from 36-site exact diagonalization of Eq. (\ref{eq:HeisenbergChiralHam}). (b) Minimum chiral spin liquid (CSL) excitation gap $\Delta_{\textrm{CSL}}$ above two-fold degenerate ground states, under flux insertion through the torus. Inset depicts the two-fold quasi-degenerate ground states (blue and dashed-red lines) and gap to excited states (brown lines), as a function of flux $\phi$ threaded through the torus [see Methods section]. The equilibrium ground state (a putative $Z_2$ spin liquid or valence bond crystal) transitions into a photo-induced CSL at finite pump strength. (c) R\'enyi entropies and entanglement minima, of CSL ground state superpositions $\cos(\theta) \ket{\psi_1} + \sin(\theta) e^{i\phi} \ket{\psi_2}$ for two bipartitions of the 36-site torus [inset], discussed in the main text.}
	\label{fig:CSL}
\end{figure}

\begin{figure}
	\centering
	\includegraphics[width=16cm]{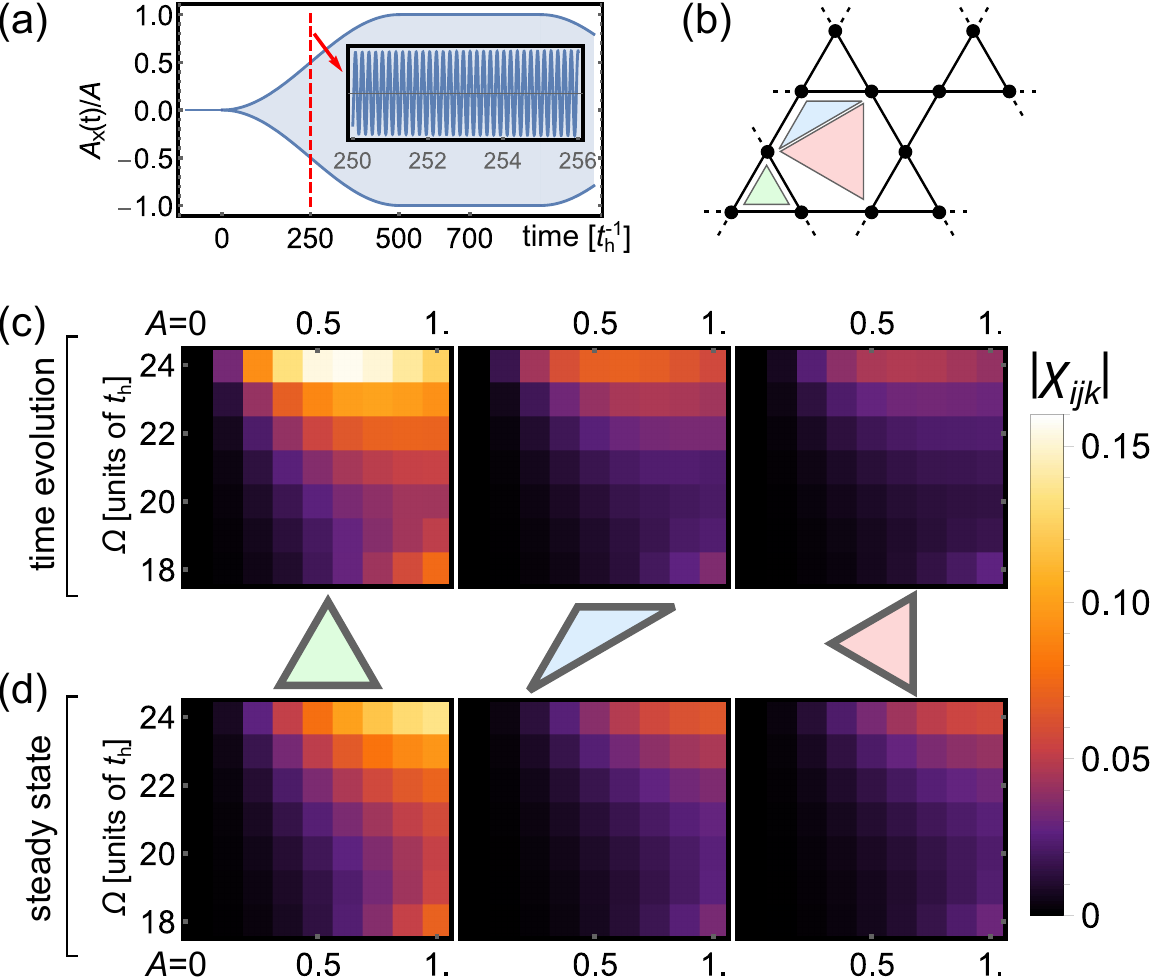}
	\caption{\textbf{Time evolution for broad-pulse irradiation.} Comparison of the Floquet chiral spin model [Eq. (\ref{eq:HeisenbergChiralHam})] and the exact time evolution of the 12-site $U=30$ Kagome Hubbard model [Eq. (\ref{eq:PumpedKagomeHubbard})] driven by circularly-polarized light with ultra-slow adiabatic pump envelopes (a). The scalar spin chiralities $\chi_{ijk} = \expect{ \S_i \cdot (\S_j \times \S_k) }$ on elementary triangles of the unit cell (b) are depicted in (c) and (d). (c)  depicts $\chi_{ijk}(t)$ measured from exact time evolution and averaged over the pump plateau; (d) depicts corresponding static expectation values of the Floquet spin model with Heisenberg and chiral couplings chosen as a function of $A,\Omega$.}
	\label{fig:HubbardVsFloquet}
\end{figure}

\begin{figure}
	\centering
	\includegraphics[width=16cm]{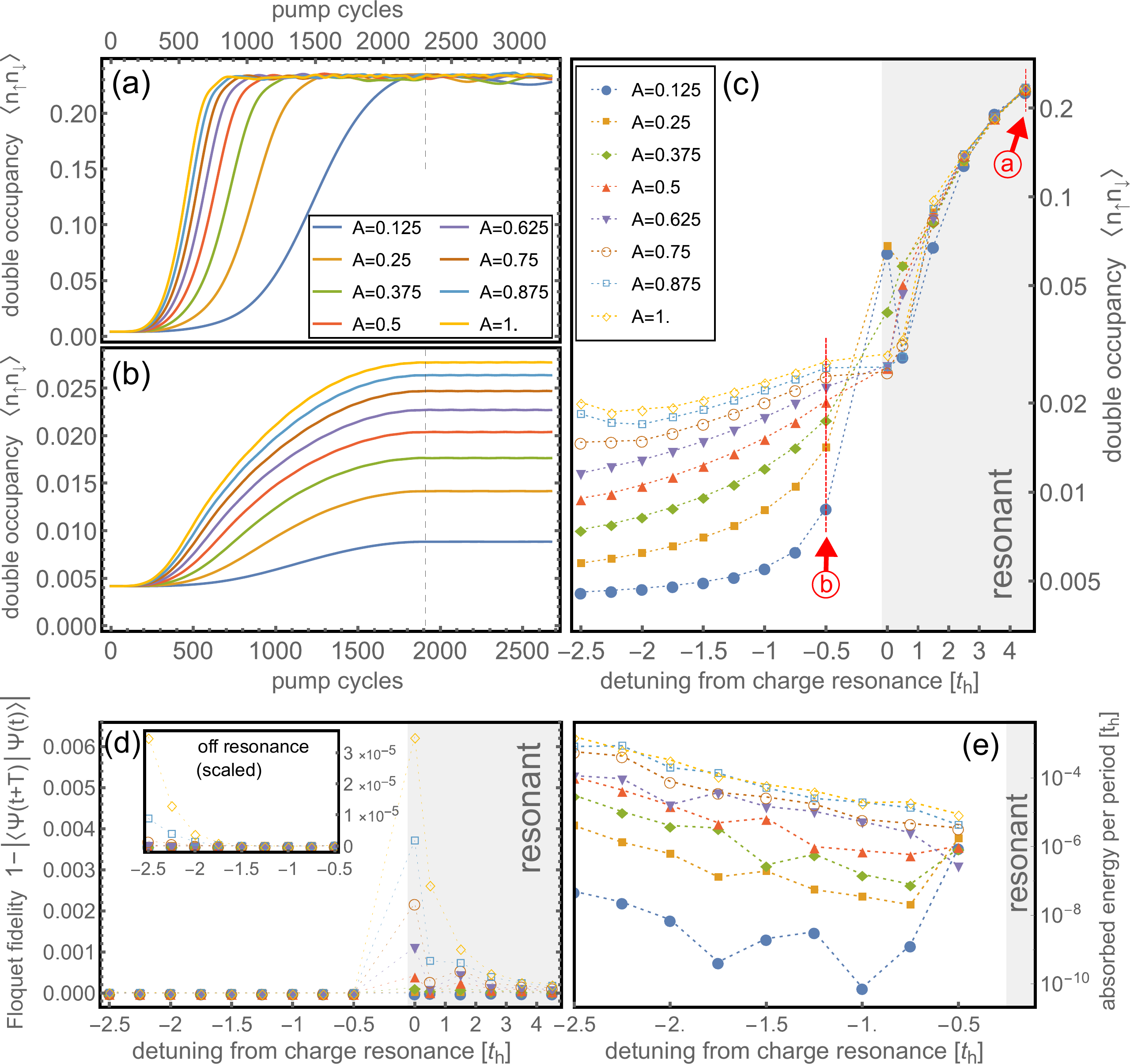}
	\caption{\textbf{Pre-thermalization and transient Floquet steady states.} (a), (b) depict the time evolution of the period-averaged local double occupancy of the 12-site driven Hubbard model as a function of cycles under the pump, on and off the charge resonance with the upper Hubbard band, respectively. Dashed lines indicate onset of the pump plateau. (c) Extracted transient expectation values of the double occupancy at the pump plateau, as a function of pump strength $A$ and detuning from the upper Hubbard band. Red arrows indicate values of $\thop$ used for (a), (b). Lines are guides to the eye. On resonance (gray region), the system heats rapidly, with $\expect{\ND{\uparrow}\ND{\downarrow}}$ thermalizating independent of the pump strength $A$ and approaching its infinite-temperature expectation value $1/4$. Below resonance (white region), the system transiently realizes the TRS-breaking chiral quantum magnet, with a tunable Hubbard interaction $U$ (as well as correspondingly tunable magnetic $J, \chi$, see Fig. \ref{fig:HubbardVsFloquet}) as a function of pump strength. (d) Floquet fidelity $F(T) = \left|\braket{\Psi(t+T)}{\Psi(t)}\right|$ on the pump plateau -- below charge resonance, $(1-F) \to 0$, indicating the controlled preparation of a Floquet eigenstate. (e) Extracted stroboscopic heating rates per pump cycle and below resonance. Remarkably, heating is strongly suppressed close to the charge resonance, with the driven system requiring many thousands of pump cycles to finally absorb energy on the order of the Heisenberg exchange $J$.}
	\label{fig:heating}
\end{figure}

The ground state degeneracy and minimal many-body excitation gap under flux insertion was calculated by imposing a spin-dependent twist of boundary conditions along one direction and tracking the many-body spectral flow as a function of twist angle, as depicted in the inset of Fig. \ref{fig:CSL}(b). A fine sampling of flux insertion, as depicted in Fig. \ref{fig:CSL}(b), was performed for 30-site clusters, spanned by vectors $\mathbf{R}_1 = 2\mathbf{a}_1 + \mathbf{a}_2, ~\mathbf{R}_2 = -2\mathbf{a}_1 + 4\mathbf{a}_2$, and checked against the 36-site cluster. The winding of the quasi-degenerate ground states upon flux insertion is a signature of CSLs, with the two quasi-degenerate ground states exchanging once under flux insertion, or remaining separated, depending on whether they lie in different (30-site cluster) or the same (36-site cluster) momentum sectors.

We furthermore consider two bipartitions $A, B$ of the 36-site cluster with periodic boundary conditions, as depicted in the inset in Fig. \ref{fig:CSL}(c), and calculate the R\'enyi entropies
\begin{align}
	S_\alpha(\theta,\phi) = -\log \textrm{tr}\left\{ \rho_{\alpha}^2(\theta,\phi) \right\}, ~~ \alpha = A,B
\end{align}
where $\rho_\alpha(\theta,\phi) = \textrm{tr}_\alpha \left\{ \ketbra{\Psi(\theta,\phi)}{\Psi(\theta,\phi)} \right\}$ is the reduced density matrix on bipartition $\alpha = A, B$ for superpositions $\ket{\Psi(\theta,\phi)} = \cos(\theta) \ket{\psi_1} + \sin(\theta) e^{i\phi} \ket{\psi_2}$ of the quasi-degenerate ground states $\ket{\psi_1}, \ket{\psi_2}$, as a function of $\theta, \phi$. Fig. \ref{fig:CSL}(c) depicts $S_A$ and $S_B$ calculated from two-fold quasi-degenerate ground states of the Floquet chiral spin model. For a CSL, $S_\alpha(\theta,\phi)$ is expected to display two entanglement minima; the two corresponding minimally-entangled states $\ket{\Psi(\theta,\phi)}$ permit extraction of the modular matrices \cite{ZhangPRB2012}, which match expectations for a Kalmeyer-Laughlin CSL and are quoted in the main text.

\subsection*{Time Evolution}

The electronic many-body time evolution was simulated for a 12-site Kagome-Hubbard cluster (4 unit cells) with periodic boundary conditions, spanned by vectors $\mathbf{R}_1 = 2\mathbf{a}_1, ~\mathbf{R}_2 = 2\mathbf{a}_2$, and with the time propagation employing adaptive step size control. We note this is the minimum cluster size to faithfully host all permutations of virtual hopping processes that give rise to the effective Floquet chiral spin Hamiltonian discussed in the main text [Eq. (\ref{eq:HeisenbergChiralHam})]. 

To model broad circularly-polarized pump pulses, we consider a pulsed field
\begin{align}
	\mathbf{A}(t) = A(t) \left[ \cos(\Omega t),~ \sin(\Omega t) \right]^\top
\end{align}
with a smooth sinusoidal pump envelope
\begin{align}
	A(t) = \left\{\begin{array}{ll}
			0,~ & t \leq 0	\\s
			A \sin^2\left( \frac{\pi}{2} \frac{t}{t_{\textrm{plateau}}} \right),~ & 0 < t < t_{\textrm{plateau}} \\
			A,~ & t \geq t_{\textrm{plateau}} \end{array} \right.
\end{align}
where $t_{\textrm{plateau}} = 700 \thop^{-1}$ for the results of the main text. Details on pump envelope dependence can be found in Supplementary Note 3.

Finally, to quantify energy absorption in the driven system, we compute the period-averaged energy operator
\begin{align}
	\hat{E} &= \frac{1}{T} \int_t^{t+T} dt' \Ham(t') \notag\\
	&= U \ND{\uparrow}\ND{\downarrow} - \BesselJ_0(A) \thop \sum_{\left<ij\right>\sigma} \CD{i\sigma}\C{j\sigma}
\end{align}
where $\BesselJ_0(A)$ denotes the zeroth Bessel function of the first kind. Note that $\hat{\expect{E}}$ is time-independent for a pure Floquet state, in theory. Instead, the finiteness of the pump envelope entails a residual quasi-energy spread, with the resulting dephasing of the driven state leading to residual heating on the pump plateau. While the driven state is ultimately expected to thermalize to an infinite-temperature state at infinite times, the results of the main text demonstrate a long-lived and remarkably stable pre-thermalized regime with negligible absorption.

\section*{Data availability}

The data that support the results presented in this study are available from the corresponding authors on request.

\newcommand{\reftitle}[1]{\textit{#1}, }
\newcommand{\refvol}[1]{\textbf{#1}}




\onecolumngrid

\pagebreak

\newcommand{\tF}[2]{t_{#1}^{(#2)}}

\clearpage
\appendix

\section{Perturbation Theory for the Non-Interacting Model}

Before discussing sub-gap pumping in the interacting Mott insulator, it is instructive to first consider electrons on the Kagome lattice in the absence of interactions, described by a Bloch Hamiltonian
\begin{align}
	h(\mathbf{k}) = -2\thop \left[\begin{array}{ccc}  0 & \cos\left(\frac{\a_1\cdot\mathbf{k}}{2}\right)  &  \cos\left(\frac{\a_2\cdot\mathbf{k}}{2}\right)  \\
	\cos\left(\frac{\a_1\cdot\mathbf{k}}{2}\right)   &   0   &   \cos\left(\frac{\a_3\cdot\mathbf{k}}{2}\right) \\
	\cos\left(\frac{\a_2\cdot\mathbf{k}}{2}\right)  &  \cos\left(\frac{\a_3\cdot\mathbf{k}}{2}\right)   &   0   \end{array}\right]
\end{align}
In equilibrium, the band structure mirrors graphene, with two Dirac points at the corners $\mathbf{K}, \mathbf{K}'$ of the Brillouin zone, and a third flat band with a quadratic band touching at $\boldsymbol{\Gamma}$.

A circularly-polarized pump field now couples to electrons via Peierls substitution $\mathbf{k} \to \mathbf{k} + \mathbf{A}(t)$, with units $\hbar = e = 1$. As there is no charge gap in the absence of interactions, we consider here the off-resonant high-frequency regime with $\Omega$ much larger than the electronic band width. At weak pump strengths, the effective Hamiltonian now follows as
\begin{align}
	h_{\textrm{eff}}(\mathbf{k}) &= h_0(\mathbf{k}) - \frac{1}{\Omega} \left[ h_1(\mathbf{k}), ~ h_{-1}(\mathbf{k}) \right] \notag\\
	&= -2 \left[\begin{array}{ccc}  0 & \tilde{t} \cos\left(\frac{\a_1\cdot\mathbf{k}}{2}\right)  &  \tilde{t}^\star \cos\left(\frac{\a_2\cdot\mathbf{k}}{2}\right)  \\
	\tilde{t}^\star  \cos\left(\frac{\a_1\cdot\mathbf{k}}{2}\right)   &   0   &   \tilde{t} \cos\left(\frac{\a_3\cdot\mathbf{k}}{2}\right) \\
	\tilde{t}  \cos\left(\frac{\a_2\cdot\mathbf{k}}{2}\right)  &  \tilde{t}^\star  \cos\left(\frac{\a_3\cdot\mathbf{k}}{2}\right)   &   0   \end{array}\right] \notag\\ &- 2i\tilde{t}' \left[\begin{array}{ccc}  0 & \cos\left(\frac{(\a_2-\a_3)\cdot\mathbf{k}}{2}\right)  &  -\cos\left(\frac{(\a_1+\a_3)\cdot\mathbf{k}}{2}\right)  \\
	-\cos\left(\frac{(\a_2-\a_3)\cdot\mathbf{k}}{2}\right)   &   0   &   \cos\left(\frac{(\a_1+\a_2)\cdot\mathbf{k}}{2}\right) \\
	\cos\left(\frac{(\a_1+\a_3)\cdot\mathbf{k}}{2}\right)  &  -\cos\left(\frac{(\a_1+\a_2)\cdot\mathbf{k}}{2}\right)   &   0   \end{array}\right]
\end{align}
with lattice vectors $\a_1 = (1,~0)^T$, $\a_2 = (1/2,~\sqrt{3}/2)^T$, and $h_m(\mathbf{k}) = \frac{\Omega}{2\pi} \int_0^{2\pi/\Omega} dt~ e^{im\Omega t} h(\mathbf{k} + \mathbf{A}(t))$. The effective Hamiltonian is now parameterized by time-reversal symmetry breaking nearest and next-nearest neighbor hoppings
\begin{align}
	\tilde{t} &= \thop \left( 1 - \frac{A^2}{4} \right) - i \frac{\sqrt{3} \thop^2 A^2}{4\Omega} \\
	\tilde{t}' &= \frac{\sqrt{3} \thop^2 A^2}{4\Omega}
\end{align}
where $A$ is the dimensionless field strength as defined in the main text. Time-reversal symmetry breaking entails that a gap opens up at the two Dirac points at $\mathbf{K}, \mathbf{K}'$, however the quadratic band touching at $\boldsymbol{\Gamma}$ persists to lowest order in $A$.

\section{Quasi-Degenerate Perturbation Theory for the Driven Kagome Hubbard Model}

Consider now a strongly-interacting Mott-Hubbard insulator on the Kagome lattice, driven by circularly-polarized light. To simplify notation, we consider a generic driven Hubbard model
\begin{align}
	\Ham(\tau) = -\sum_{ij\sigma} t_{ij}(\tau)~ \CD{i\sigma}\C{j\sigma} + U \sum_i \ND{i\uparrow}\ND{i\downarrow}  \label{eq:PumpedKagomeHubbard2}
\end{align}
where the external drive enters in the time-dependent hopping amplitudes, and we denote time by $\tau$. The case of external drive via Peierls substitution corresponds to $t_{ij}(\tau) = \thop e^{i \mathbf{A}(\tau) \cdot \mathbf{r}_{ij}}$ as discussed in the main text. In Floquet language, one can similarly write:
\begin{align}
	\Ham &= -\sum_{\substack{ij\sigma \\ mm'}} t_{ij}^{(m-n)} \CD{i\sigma}\C{j\sigma} \otimes \ketbra{m}{m'} +  U \sum_{i} \ND{i\uparrow}\ND{i\downarrow} \otimes \mathbf{1} - \sum_{m} m\Omega ~\mathbf{1} \otimes \ketbra{m}{m} \label{eq:HamFloquetHubbard2}
\end{align}
where $t_{ij}^{(m-n)} = \int_0^{2\pi/\Omega} d\tau~ e^{i(m-n)\tau} t_{ij}(\tau)$ are Floquet hopping amplitudes that obey $\left( t_{ji}^{(-m)} \right)^\star = t_{ij}^{(m)}$.

So far, this description is exact. As described in the main text, the Hamiltonian retains a local moment description of interacting spins as long as $U$ is much larger than hopping, and the pump is off-resonant with charge excitations. A spin description for pumping below the charge gap then follows from \textit{simultaneously} integrating out doubly-occupied many-body states and photon (Floquet) side bands, treating strong-coupling and Floquet energy scales on equal footing.

Before discussing the derivation in detail, it is instructive to compare our methodology to the Schrieffer-Wolff transformation for driven systems \cite{BukovPRL2016S}. The methods are \textit{a priori} equivalent for pumping off the charge resonance (the regime of interest in this work), which is to be expected, as the perturbation expansion should give the same results independent of choice of basis. Importantly however, as discussed in detail below, the lowest-order contribution in virtual hopping merely serves to ``renormalize'' the strength of nearest-neighbor spin exchange interactions \cite{BukovPRL2016S,MentinkNComm2015S}, which \textit{cannot change the transient many-body state} as long as total spin is conserved. Instead, as discussed in the main text, the subleading photon-assisted contributions -- to fourth order in virtual hopping -- are essential to manipulate the equilibrium state, break time-reversal symmetry, and ultimately stabilize a chiral spin liquid.

\subsection*{Second Order in Virtual Hopping}

In Floquet language, simultaneously integrating out charge and Floquet degrees of freedom as discussed above amounts to applying standard quasi-degenerate perturbation theory to the time-independent Floquet Hamiltonian in frequency formulation. We constrain ourselves to the half-filled Mott insulator. To second order in virtual hopping processes, the perturbation theory entails a renormalization of nearest-neighbor Heisenberg exchange interactions
\begin{align}
	\Ham^{(\textrm{2nd order})} &= \sum_{ij\sigma\sigma'} \frac{\tF{ij}{-m}\tF{ji}{m}}{U + m\Omega}~  \CD{i\sigma}\C{j\sigma} \CD{j\sigma'}\C{i\sigma'} \notag\\
	&= 2\sum_{\left<ij\right>\sigma\sigma'} \frac{\left| \tF{ji}{m} \right|^2}{U + m\Omega}~ \CD{i\sigma}\C{i\sigma'}\C{j\sigma}\CD{j\sigma'} \notag\\
	&= 4\sum_{\left<ij\right>} \frac{\left| \tF{ji}{m} \right|^2}{U + m\Omega}~ \mathbf{S}_i \cdot \mathbf{S}_j
\end{align}
where the last line follows from a canonical change to spin-1/2 operators and the local moment constraint. Choosing Floquet hopping amplitudes $\tF{ij}{m}$ that correspond to circularly-polarized light finally leads to the result quoted in the main text.


\subsection*{Third Order in Virtual Hopping}

As a short digression, a more interesting scenario appears to third-order in virtual hopping. In equilibrium, it is well-known that the third-order expansion vanishes in the presence of time-reversal symmetry. Conversely, if time-reversal symmetry is broken via an external magnetic field, a scalar spin chirality term appears already to this order, and is proportional to the phase acquired by electrons hopping around a triangle. Na\"ively, a similar scenario should then arise already to third order, in the presence of a circularly-polarized field, with electrons acquiring a phase when absorbing or emitting photons during virtual hopping. Here, we show that this is not the case.

To third order in virtual hopping, and without any assumptions on symmetries of the problem, the generic fermionic starting point of the expansion reads
\begin{align}
	\Ham^{(3)} = \sum_{\substack{ijk \\ \sigma\sigma'\sigma'' \\ m_1m_2}} \frac{1}{(U+m_1\Omega)(U+m_2\Omega)} &\left[ \CD{i\sigma}\C{j\sigma}\CD{j\sigma'}\C{k\sigma'}\CD{k\sigma''}\C{i\sigma''} \tF{ij}{-m_2} \tF{jk}{m_2-m_1} \tF{ki}{m_1} + \right. \notag\\
	+& \left.  ~\CD{j\sigma'}\C{k\sigma'}\CD{i\sigma}\C{j\sigma}\CD{k\sigma''}\C{i\sigma''} \tF{jk}{-m_2} \tF{ij}{m_2-m_1} \tF{ki}{m_1} \right] \notag\\
\end{align}
where $i, j, k$ sum over all sites. After appropriate expansion of all permutations, reordering of fermionic operators and recasting in terms of spin-1/2 operators, one arrives at a Heisenberg-chiral spin Hamiltonian\begin{align}
	\Ham^{\textrm{3rd order}} = \sum_{\left<ij\right>} J^{\textrm{(3rd order)}} ~\mathbf{S}_i \cdot \mathbf{S}_j + \sum_{\substack{\bigtriangleup \\ ijk}} \chi^{\textrm{(3rd order)}} ~\mathbf{S}_i \cdot \left( \mathbf{S}_j \times \mathbf{S}_k \right)
\end{align}
Here, the Heisenberg exchange coupling and scalar spin chirality coupling can be expressed as
\begin{align}
	\chi^{\textrm{3rd order}} &= \sum_{m_1m_2}\frac{2}{(U+m_1\Omega)(U+m_2\Omega)} \textrm{Im} \left[  \tF{ij}{m_2} \tF{jk}{m_1-m_2} \tF{ki}{-m_1} + \tF{ij}{m_1-m_2} \tF{jk}{m_2} \tF{ki}{-m_1} +  \right. \notag\\
&+ \tF{ij}{m_2-m_1} \tF{jk}{-m_2} \tF{ki}{m_1} + \tF{ij}{-m_2} \tF{jk}{m_2-m_1} \tF{ki}{m_1} + \tF{ij}{m_2} \tF{jk}{-m_1} \tF{ki}{m_1-m_2} + \tF{ij}{-m_1} \tF{jk}{m_2} \tF{ki}{m_1-m_2} + \notag\\
&+ \tF{ij}{m_2-m_1} \tF{jk}{m_1} \tF{ki}{-m_2} + \tF{ij}{m_1} \tF{jk}{m_2-m_1} \tF{ki}{-m_2} + \tF{ij}{m_1-m_2} \tF{jk}{-m_1} \tF{ki}{m_2} + \tF{ij}{-m_1} \tF{jk}{m_1-m_2} \tF{ki}{m_2}  + \notag\\
&+ \tF{ij}{-m_2} \tF{jk}{m_1} \tF{ki}{m_2-m_1} + \left. \tF{ij}{m_1} \tF{jk}{-m_2} \tF{ki}{m_2-m_1}  \right]  \label{eq:chiralThirdOrder} \\
\
	J^{\textrm{3rd order}} &= \sum_{m_1m_2}\frac{1}{(U+m_1\Omega)(U+m_2\Omega)} \textrm{Re}\left[-(\tF{ij}{m_2} \tF{jk}{m_1-m_2} \tF{ki}{-m_1}) - \tF{ij}{m_1-m_2} \tF{jk}{m_2} \tF{ki}{-m_1} + \right.\notag\\
&+ \tF{ij}{m_2-m_1} \tF{jk}{-m_2} \tF{ki}{m_1} + \tF{ij}{-m_2} \tF{jk}{m_2-m_1} \tF{ki}{m_1} + \tF{ij}{m_2} \tF{jk}{-m_1} \tF{ki}{m_1-m_2} - \tF{ij}{-m_1} \tF{jk}{m_2} \tF{ki}{m_1-m_2} -  \notag\\
&- \tF{ij}{m_2-m_1} \tF{jk}{m_1} \tF{ki}{-m_2} - \tF{ij}{m_1} \tF{jk}{m_2-m_1} \tF{ki}{-m_2} + \tF{ij}{m_1-m_2} \tF{jk}{-m_1} \tF{ki}{m_2} + \tF{ij}{-m_1} \tF{jk}{m_1-m_2} \tF{ki}{m_2} -  \notag\\
&\tF{ij}{-m_2} \tF{jk}{m_1} \tF{ki}{m_2-m_1} + \tF{ij}{m_1} \tF{jk}{-m_2} \tF{ki}{m_2-m_1} \left.\right]   \label{eq:heisenbergThirdOrder}
\end{align}
To proceed, consider a generic parameterization of the Floquet hoppings
\begin{align}
	\tF{ij}{m} = t_{ij} \BesselJ_{m}(A_{ij}) e^{i m \psi_{ij}}
\end{align}
Importantly, one then finds that the terms in the bracket in Supplementary Eq. (\ref{eq:chiralThirdOrder}) are real and the chiral contribution $\chi^{\textrm{3rd order}}$ \textit{vanishes exactly even if time-reversal symmetry is broken} is by the external field. In other words, the photon-mediated phases acquired by electrons hopping around a triangle remarkably cancel exactly and to all orders in pump strength $A$. The third-order Heisenberg contribution $J^{\textrm{3rd order}}$ vanishes analogously, with the terms in the bracket in Supplementary Eq. (\ref{eq:heisenbergThirdOrder}) purely imaginary.

Instead, in analogy to the equilibrium case, solely a \textit{static} magnetic field will induce a scalar spin chirality term to third order. As discussed in the main text however, in this case the Zeeman shift will generically dominate for an external magnetic field and preclude the formation of a chiral spin liquid.

For completeness, we could now consider the joint effect of a \textit{static} magnetic field and time-dependent circularly-polarized pump. Here, the nearest-neighbor Floquet hopping amplitudes
\begin{align}
	\tF{\left<ij\right>}{m} = e^{i\phi_{\textrm{mag}} } ~ \thop \BesselJ_{m}(A_{ij}) e^{i m \psi_{ij}}
\end{align}
would acquire an additional phase $\phi_{\textrm{mag}}$ due to the static magnetic flux through the triangle. Unlike the equilibrium problem, the third-order contribution to the effective Hamiltonian in this case entails not only a scalar spin chirality term but further corrections to nearest-neighbor Heisenberg exchange
\begin{align}
	J^{\textrm{3rd order}} &= \sum_{m_1m_2} \frac{4 \thop^3 \BesselJ_{m_1}(A) \BesselJ_{m_2-m_1}(A) \BesselJ_{-m_2}(A) }{(U+m_1\Omega)(U+m_2\Omega)} \sin\left[\tfrac{2\pi(2m_1-m_2)}{3} \right] ~ \sin(3\phi_{\textrm{mag}}) \\
	\chi^{\textrm{3rd order}} &= \sum_{m_1m_2} \frac{24 \thop^3 \BesselJ_{m_1}(A) \BesselJ_{m_2-m_1}(A) \BesselJ_{-m_2}(A) }{(U+m_1\Omega)(U+m_2\Omega)} \cos\left[\tfrac{2\pi(2m_1-m_2)}{3}\right]  ~ \sin(3\phi_{\textrm{mag}})
\end{align}
The well-known equilibrium result \cite{MacDonaldPRB1988,SenPRB1995,MotrunichPRB2006} for solely a static magnetic field can be readily recovered setting $A$ to zero, which yields $J^{\textrm{3rd order}} = 0$ and $\chi^{\textrm{3rd order}} = 24 \thop^3 / U^2 \sin(3\phi_{\textrm{mag}})$ .


\subsection*{Fourth Order in Virtual Hopping}

Constituting the central result of the main text, dynamical time-reversal symmetry breaking due to circularly-polarized light first enters the effective spin description to fourth order in photon-assisted virtual hopping processes. The presence and relevance of this effect in determining the transient steady state is remarkable -- whereas fourth-order contributions to spin interactions will commonly be negligible in equilibrium Mott insulators, the out-of-equilibrium problem provides a powerful knob to tune the spin system into a regime where such contributions emerge, in fact, as the determining mechanism to stabilize the transient phase.

The quasi-degenerate perturbation theory again proceeds as usual, however entails a large number of combinatoric contributions due to the various permutations of hopping virtually through different Floquet side bands. We therefore developed a computer algebra program to perform the perturbation expansion and recasting into spin language. Here, we solely quote the fermionic starting point of the fourth-order expansion
\clearpage
\begin{align}
	\Ham^{(4)} = - \sum_{\substack{ijkl \\ \sigma\sigma'\sigma''\sigma''' \\ m_1m_2m_3}} &\left[ \CD{i\sigma}\C{j\sigma} \CD{j\sigma'}\C{k\sigma'} \CD{k\sigma''}\C{l\sigma''} \CD{l\sigma'''}\C{i\sigma'''}   \frac{(1-\delta_{ik}\delta_{m_2}) ~~ \tF{ij}{-m_3} \tF{jk}{m_3-m_2} \tF{kl}{m_2-m_1} \tF{li}{m_1} }{(U+m_1\Omega)(U(1-\delta_{ik})+m_2\Omega)(U+m_3\Omega)} + \right. \notag\\
\
	&+ \CD{j\sigma'}\C{k\sigma'} \CD{i\sigma}\C{j\sigma} \CD{k\sigma''}\C{l\sigma''} \CD{l\sigma'''}\C{i\sigma'''}  \frac{(1-\delta_{ik}\delta_{m_2}) ~~ \tF{jk}{-m_3} \tF{ij}{m_3-m_2} \tF{kl}{m_2-m_1} \tF{li}{m_1} }{(U+m_1\Omega)(U(1-\delta_{ik})+m_2\Omega)(U+m_3\Omega)} + \notag\\
\
	&+ \CD{j\sigma'}\C{k\sigma'} \CD{k\sigma''}\C{l\sigma''} \CD{i\sigma}\C{j\sigma} \CD{l\sigma'''}\C{i\sigma'''}  \frac{(1-\delta_{jl}\delta_{m_2}) ~~ \tF{jk}{-m_3} \tF{kl}{m_3-m_2} \tF{ij}{m_2-m_1} \tF{li}{m_1} }{(U+m_1\Omega)(U(1-\delta_{jl})+m_2\Omega)(U+m_3\Omega)} + \notag\\
	&+ \left. \CD{k\sigma''}\C{l\sigma''} \CD{j\sigma'}\C{k\sigma'} \CD{i\sigma}\C{j\sigma} \CD{l\sigma'''}\C{i\sigma'''}  \frac{(1-\delta_{jl}\delta_{m_2}) ~~ \tF{kl}{-m_3} \tF{jk}{m_3-m_2} \tF{ij}{m_2-m_1} \tF{li}{m_1} }{(U+m_1\Omega)(U(1-\delta_{jl})+m_2\Omega)(U+m_3\Omega)} \right] + \notag\\
\
	-~ \sum_{\substack{ijkl \\ \sigma\sigma'\sigma''\sigma''' \\ m_1m_2m_3}} & \frac{1}{(U+m_1\Omega)(2U+m_2\Omega)(U+m_3\Omega)} \left[\vphantom{\frac{1}{2}}\right. \notag \\
	&\hspace{3.15cm} \CD{i\sigma}\C{l\sigma} \CD{j\sigma'}\C{k\sigma'} \CD{k\sigma''}\C{j\sigma''} \CD{l\sigma'''}\C{i\sigma'''} \tF{il}{-m_3} \tF{jk}{m_3-m_2} \tF{kj}{m_2-m_1} \tF{li}{m_1}  +  \notag\\
	&\hspace{3.15cm}+ \CD{j\sigma}\C{l\sigma} \CD{i\sigma'}\C{k\sigma'} \CD{k\sigma''}\C{j\sigma''} \CD{l\sigma'''}\C{i\sigma'''} \tF{jl}{-m_3} \tF{ik}{m_3-m_2} \tF{kj}{m_2-m_1} \tF{li}{m_1} \notag\\
	&\hspace{3.15cm}+ \CD{j\sigma}\C{k\sigma} \CD{i\sigma'}\C{l\sigma'} \CD{k\sigma''}\C{j\sigma''} \CD{l\sigma'''}\C{i\sigma'''} \tF{jk}{-m_3} \tF{il}{m_3-m_2} \tF{kj}{m_2-m_1} \tF{li}{m_1} \notag\\
	&\hspace{3.15cm}+ \left. \CD{i\sigma}\C{k\sigma} \CD{j\sigma'}\C{l\sigma'} \CD{k\sigma''}\C{j\sigma''} \CD{l\sigma'''}\C{i\sigma'''} \tF{ik}{-m_3} \tF{jl}{m_3-m_2} \tF{kj}{m_2-m_1} \tF{li}{m_1} \right]  \notag\\
\
	+~ \frac{1}{2} \sum_{\substack{ijkl \\ \sigma\sigma'\sigma''\sigma''' \\ m_1m_2}} & \CD{k\sigma}\C{l\sigma} \CD{l\sigma'}\C{k\sigma'} \CD{i\sigma''}\C{j\sigma''} \CD{j\sigma'''}\C{i\sigma'''} ~ \tF{kl}{-m_2} \tF{lk}{m_2} \tF{ij}{-m_1} \tF{ji}{m_1} \left[ \frac{1}{(U+m_1\Omega)(U+m_2\Omega)^2} ~+ \right. \notag\\
	&\hspace{6.2cm}+ \left. \frac{1}{(U+m_1\Omega)^2(U+m_2\Omega)} \right]
\end{align}
where we imply within the terms of the first sum that $\frac{(1-\delta_{ik}\delta_{m_2})}{U(1-\delta_{ik}) + m_2\Omega} = 0$ if $i=k$ and $m_2 = 0$. The resulting spin Hamiltonian for circularly-polarized pumping can now be derived from tedious but straightforward algebraic operations, and is presented in the main text.

We note that similar perturbative expansions must be used to recast electronic experimental observables in terms of the effective low-energy spin degrees of freedom. A detailed discussion of optical and Raman probes (in particular, the A$_\textrm{2g}$ channel and its analogue for the Kagome lattice) to fourth order in virtual hopping processes will be discussed in an upcoming publication.

\section{Pump-Envelope Dependence and Many-Body Dynamics of the Driven Kagome-Hubbard Model}

\begin{figure}[h]
	\subfloat[Period-averaged scalar spin chirality expectation values.]{
	\includegraphics[width=17cm,trim=2cm 1.4cm 0 0,clip]{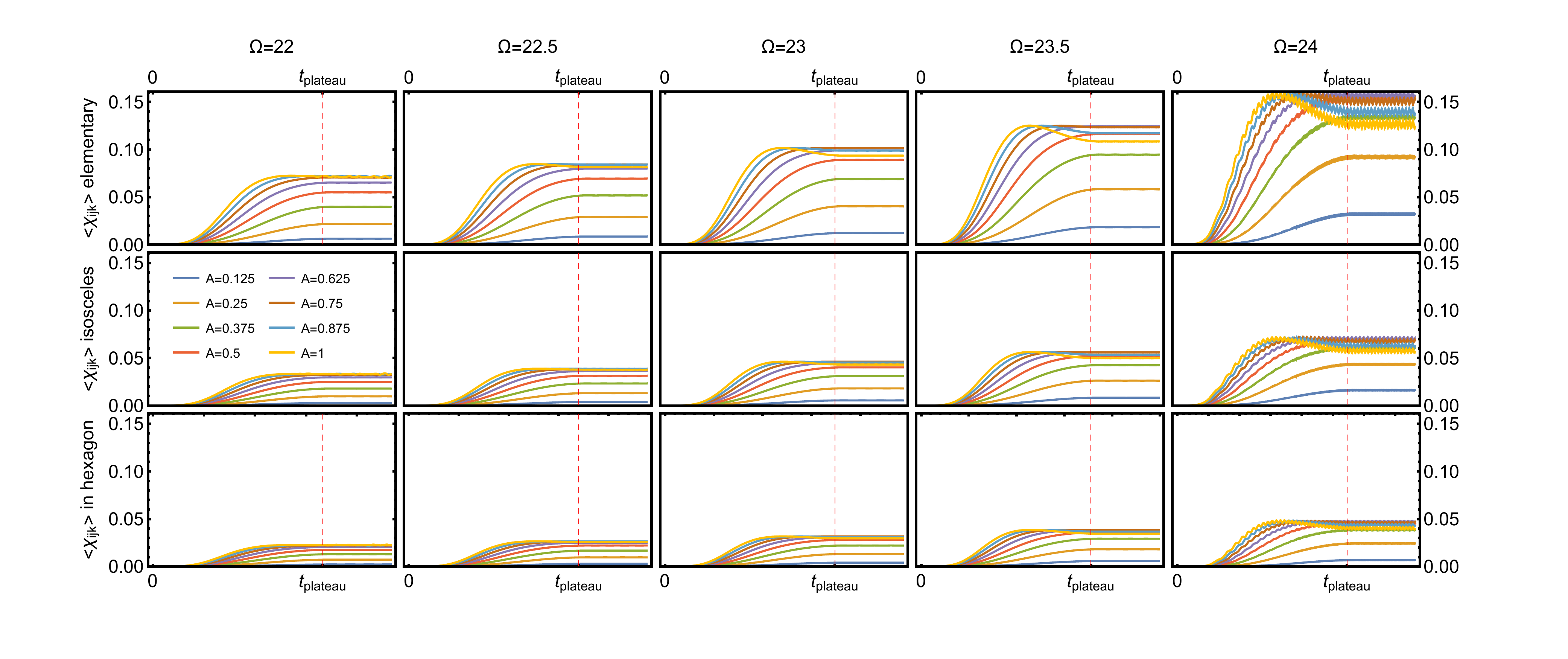}
	}
	\hfill
	\subfloat[Raw dynamical data.]{
		\includegraphics[width=16cm,trim=2cm 0 0 0,clip]{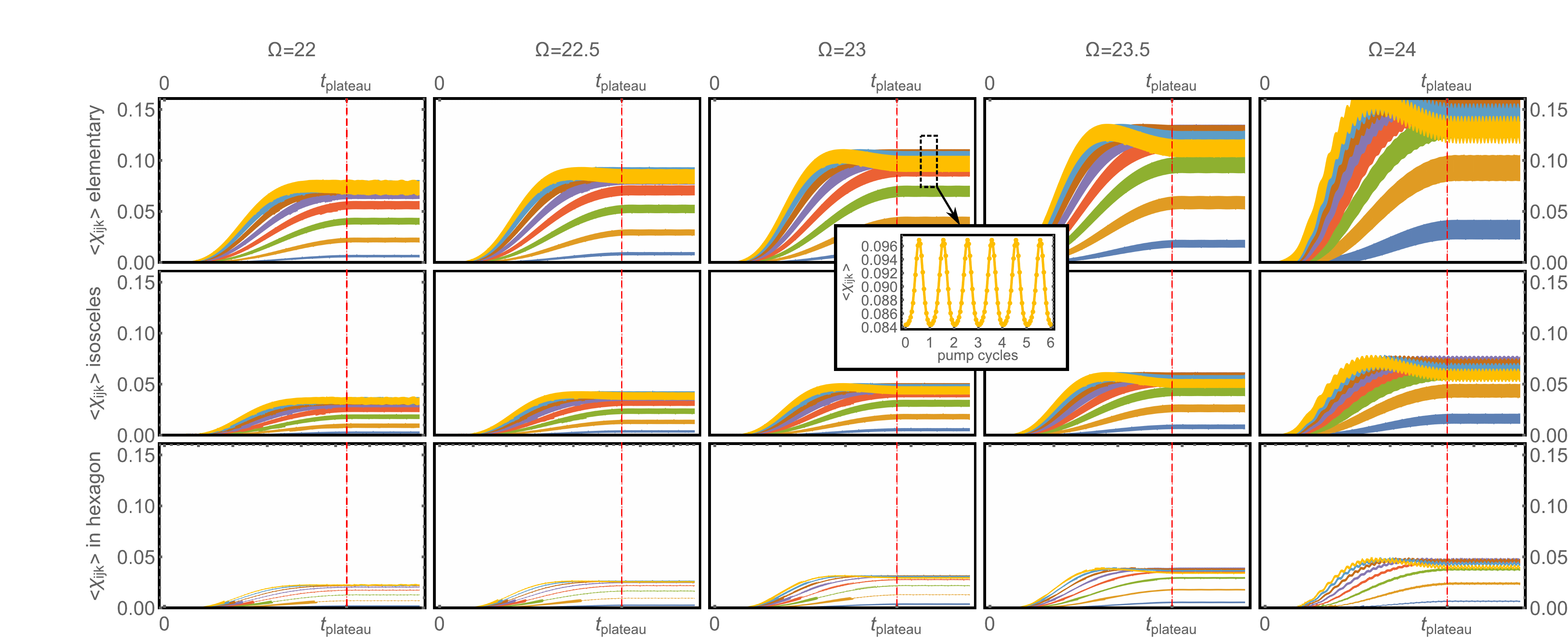}
   }
   \caption{\textbf{Time evolution of scalar spin chirality expectation values.} Panels depict time-dependent expectation values $\expect{\mathbf{S}_i \cdot \left( \mathbf{S}_j \times \mathbf{S}_k \right) (t)}$ for circularly-polarized pumping close to the one-photon charge resonance, as a function of pump strength and frequency. (a) depicts a triplet of rows corresponding to measuring the period-averaged scalar spin chirality on three triangles in the unit cell: elementary triangles of the Kagome lattice, isosceles triangles inside the hexagon, and equilateral triangles inside the hexagon, as depicted graphically in Fig. 4 (b) of the main text. (b) depicts the corresponding raw dynamical data. The broadening of lines stems from micro-motion within individual pump periods, shown in detail in the inset. Floquet expectation values are recovered by averaged out the micro-motion, as shown in (a).}
	\label{fig:Sdynamicaldata}
\end{figure}

The main text investigates the many-body dynamics of a Kagome-Hubbard model pumped with circularly-polarized light, and compares the driven steady state at the pump plateau to the effective Floquet chiral spin model that is the main focus of this work. To extract scalar spin chirality expectation values from the driven steady state at the pump plateau, we consider pump fields of the form
\begin{align}
	\mathbf{A}(t) = A(t) \left[\begin{array}{l} \cos(\Omega t) \\ \sin(\Omega t) \end{array}\right]
\end{align}
with a smooth sinusoidal pump envelope
\begin{align}
	A(t) = \left\{\begin{array}{ll}
			0,~ & t \leq 0	\\
			A \sin^2\left( \frac{\pi}{2} \frac{t}{t_{\textrm{plateau}}} \right),~ & 0 < t < t_{\textrm{plateau}} \\
			A,~ & t \geq t_{\textrm{plateau}} \end{array} \right.
\end{align}
with $t_{\textrm{plateau}} = 500 \thop^{-1}$. Supplementary Fig. \ref{fig:Sdynamicaldata} depicts the period-averaged scalar spin chirality expectation values $\expect{\chi_{ijk}}$ for three distinct triangles of the Kagome unit cell, as well as a the raw dynamical data (including micro-motion within the pump periods), close to the charge resonance for a 12-site cluster. Fig. 4(c) of the main text now follows straight-forwardly via extracting the average expectation values $\expect{\chi_{ijk}}$ at the pump plateau.

\begin{figure}[h]
	\includegraphics[width=14cm]{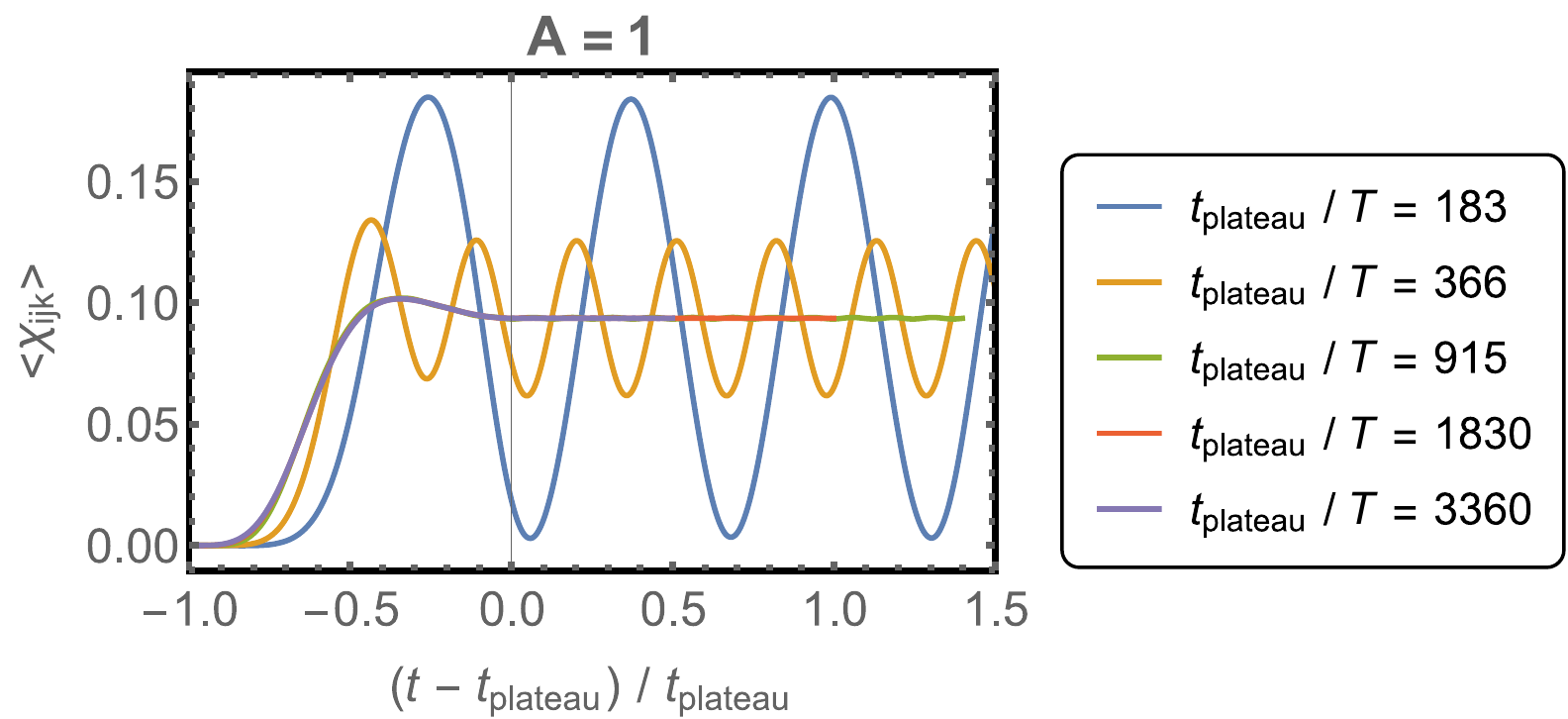}
	\caption{\textbf{Pump envelope dependence of scalar spin chirality expectation values}. The time evolution of period-averaged scalar spin chirality expectation values is shown for elementary triangles of the Kagome lattice, for $\Omega=23 \thop$. Here, $t_{\textrm{plateau}} / T$ quantifies the number of pump cycles under the ramp-up, before reaching the pump plateau at time $t = t_{\textrm{plateau}}$ (time axis is normalized accordingly). Pump envelopes used for Fig. 4 and Fig. 5 of the main text are depicted in green ($t_{\textrm{plateau}} / T = 915$).}
	\label{fig:SenvelopeChiral}
\end{figure}

\begin{figure}[h]
	\includegraphics[width=16cm,trim=1.5cm 0 0 0,clip]{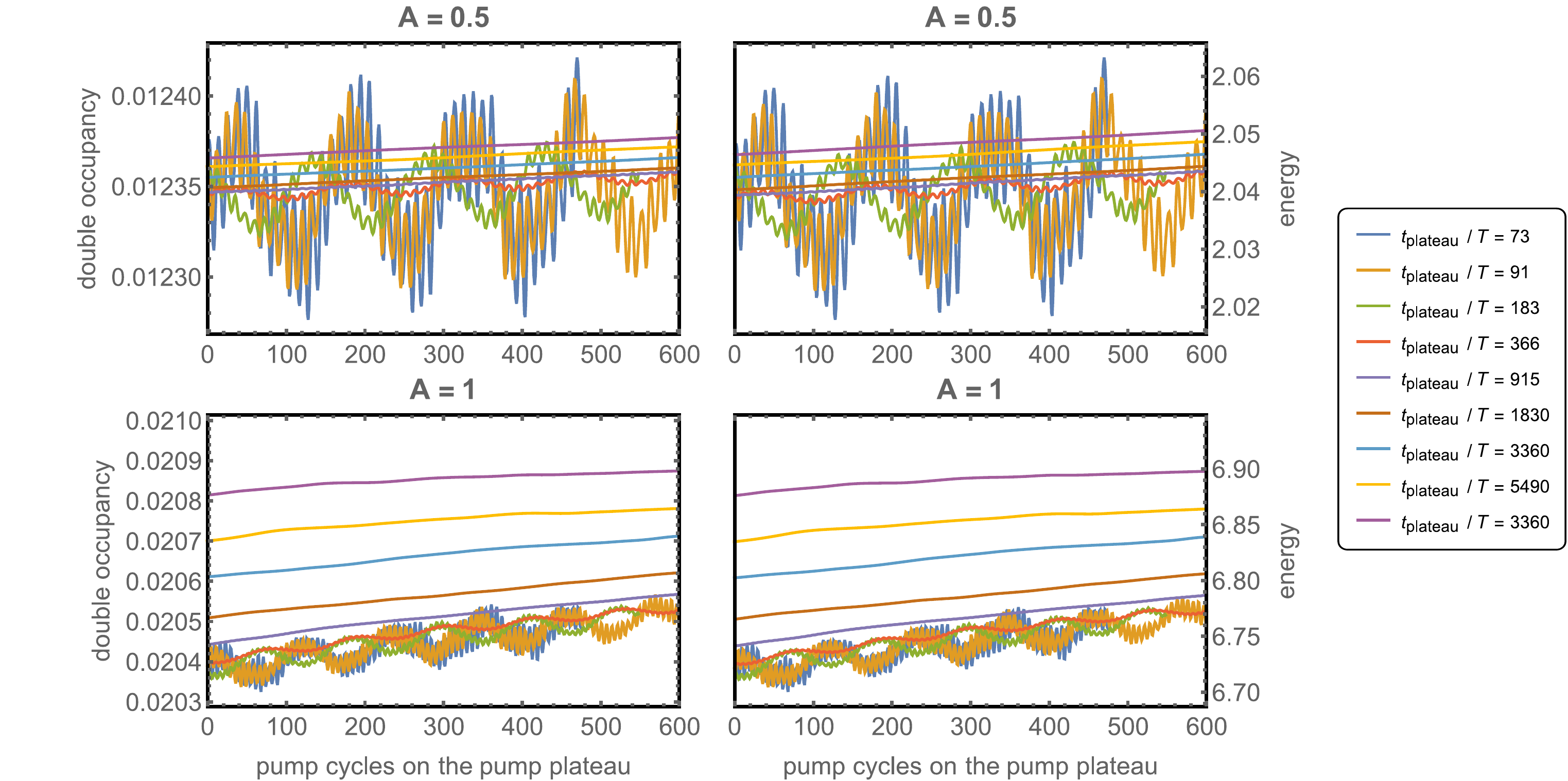}
	\caption{\textbf{Pump envelope dependence of period-averaged double occupancies and energy.} Tow (bottom) rows correspond to intermedite (strong) pumping, for $\Omega=23\thop$, on the pump plateau (pump cycle 0 $\equiv t_{\textrm{plateau}}$). Energy depicted in units of hopping strength. The time derivative of double occupancies and energy yields effective heating rates, demonstrating negligible heating both during the pump ramp-up and on the pump plateau, for pumping below the one-photon charge resonance.}
	\label{fig:SenvelopeHeating}
\end{figure}

To analyze the influence of the pump envelope on controlled preparation of the steady state, we calculate the scalar spin chirality expectation values as well as double occupancies and period-averaged energy, as a function of $t_{\textrm{plateau}}$, i.e. as a function of ramp-up speed. Supplementary Fig. \ref{fig:SenvelopeChiral} depicts the envelope dependence for scalar spin chirality expectation values, showing excellent control of the transient steady state for the pump envelope widths used in the main text. To consider residual effects of heating and energy absorption, Supplementary Fig. \ref{fig:SenvelopeHeating} depicts transient double occupancies and period-averaged energy on the pump plateau, as a function of pump envelope width, for intermediate and strong pumping. Analogously, the steady state is remarkably robust, with heating strongly suppressed. To quantify the latter, residual energy absorption rates can be extracted from the gradient of the period-averaged energy, and are depicted in Fig. 5(e) of the main text.

\clearpage

\renewcommand\refname{Supplementary References}
\def\bibsection{\section*{\refname}}

\end{document}